\documentclass[12pt,titlepage]{article}
\usepackage{amsmath}
\usepackage{amssymb}
\usepackage{graphicx}
\usepackage{caption2}
\usepackage{amsfonts}
\usepackage{cite}

\newcommand{\be}{\begin{equation}}
\newcommand{\ee}{\end{equation}}
\newcommand{\bea}{\begin{eqnarray}}
\newcommand{\eea}{\end{eqnarray}}
\newcommand{\bef}{\begin{figure}}
\newcommand{\ef}{\end{figure}}
\newcommand{\bt}{\begin{tabular}}
\newcommand{\et}{\end{tabular}}
\newcommand{\bno}{\begin{enumerate}}
\newcommand{\eno}{\end{enumerate}}

\def\3{\ss}

\catcode`\"=\active
\def"{\accent'177}
\begin{document}

\begin{center}

{\bf \Large  Physical limit of prediction \\ for chaotic motion of three-body problem }

\vspace{0.3cm}

Shijun  Liao $^{a,b,c}$

\vspace{0.3cm}

$^a$
State Key Laboratory of Ocean Engineering, Shanghai 200240, China

\vspace{0.25cm}
$^b$
School of Naval Architecture, Ocean and Civil Engineering\\
Shanghai Jiao Tong University, Shanghai 200240, China\\

\vspace{0.25cm}

$^c$ Nonlinear Analysis and Applied Mathematics Research Group (NAAM) \\  King Abdulaziz University (KAU), Jeddah, Saudi Arabia

\end{center}

\hspace{-0.5cm}{\bf Abstract}
{\em   A half century ago, Lorenz found the  ``butterfly effect'' of chaotic dynamic systems and made his famous claim that  long-term prediction of chaos is impossible.  However,  the meaning of the ``long-term'' in his claim is not very clear.   In this article,  a new concept, i.e. the physical limit of prediction time, denoted by $T_p^{max}$,  is put forwarded to provide us a time-scale for at most how long   mathematically reliable (numerical) simulations of trajectories of a chaotic dynamic system are  physically  correct.  A special case of three-body problem is used as an example to illustrate that, due to the inherent, physical uncertainty of initial positions in the (dimensionless) micro-level of $10^{-60}$,  the  chaotic trajectories  are  essentially uncertain in physics after $t > T_p^{max}$, where $T_p^{max}\approx 810$ for this special case of the three body problem.  Thus, physically,  it has no sense to talk about the ``accurate,  deterministic  prediction''  of  chaotic trajectories of the three body problem after $t > T_p^{max}$.  In addition, our mathematically reliable  simulations of the chaotic trajectories of the three bodies   suggest  that, due to the butterfly effect of chaotic dynamic systems,  the micro-level physical uncertainty of initial conditions might transfer into macroscopic uncertainty.     This suggests that  micro-level uncertainty might be an origin of some macroscopic uncertainty.  Besides, it might provide us a theoretical explanation about the origin of uncertainty  (or randomness)  of  many macroscopic phenomena such as  turbulent flows, the random distribution of stars in the universe, and so on. }

\hspace{-0.5cm}{\bf Key Words}   Three-body problem,   chaos,  physical uncertainty, multiple scales

\section{Introduction}

Why do stars in the sky look like random?   Why are velocities of turbulent flows so uncertain?  Are there any relationships  between microscopic uncertainty and macroscopic uncertainty?  What is the origin of the macroscopic uncertainty and/or randomness?    Can we give a  theoretical explanation for such kind of macroscopic uncertainty and/or randomness?   In this paper, we attempt to give such an explanation using chaotic motion of Newtonian three-body problem as an example.

Making mathematically reliable and physically correct prediction is an important scientific object.    Laplace claimed that the future can be unerringly predicted, given sufficient knowledge of the present.   However,
 Laplace was wrong even in a classical, non-chaotic universe, as pointed out by Wolpert  \cite{Wolpert2008}.   In 1890,  Poincar\'{e} \cite{Poincare1890} found that trajectories of three-body problem are unintegrable in general.   In 1963 Lorenz \cite{Lorenz1963}  found  that  it  is  impossible  to  make  ``long-term''  prediction  of  non-periodic (called chaotic today)  solution of a nonlinear dynamic system (called today Lorenz equation), due to the sensitive dependence on initial condition (SDIC), i.e.  a very slight variation of initial conditions leads to considerably obvious difference of  computer-generated trajectories for a long time.   Besides, it was found \cite{Lorenz1989, Lorenz2006, Li2000, Li2001, Teixeira2007, Liao2009, Liao2012} that  numerical simulations of chaotic trajectories are sensitive not only to initial conditions but also to (traditional) numerical algorithms and digit precision of numerical data, i.e. a slight difference of (traditional) numerical algorithms (such as a tiny variation of time step) and/or digit precision of data  might lead to considerably large difference of simulations of chaotic trajectories for a long time.    This is easy to understand, since numerical noises, i.e. truncation and round-off errors, are  inherent and unavoidable for all numerical simulations,  where truncation error is closely related to numerical algorithms and round-off error is due to the limited precision of numerical data.  According to the Shadow Lemma \cite{Anosov1967},  for a uniformly hyperbolic dynamic system, there always exists a true trajectory near any computer-generated trajectories, as long as the truncation and round-off errors are small enough.  Unfortunately, hardly a nonlinear dynamic system is  uniformly hyperbolic in most cases so that the  Shadow Lemma \cite{Anosov1967}  often does not wok in practice.   Besides,  it is found that  for some chaotic dynamic systems,   computer-generated trajectories can be shadowed only for a short time \cite{Yorke1994},  and in addition  it is ``virtually impossible to obtain a long trajectory that is even approximately correct'' \cite{Yorke1997}.  Furthermore,   For some chaotic systems,  ``there is no fundamental reasons for computer-simulated long-time statistics to be even approximately correct'' \cite{Sauer2002}.   As illustrated by Yuan and York \cite{Yorke2000} using a model, a numerical artifact persists for an arbitrarily high numerical precision.   These numerical artifacts ``expose an exigent demand of safe numerical simulations''  \cite{Shi2008}.  Thus,  it is a huge challenge to give mathematically reliable simulations of chaotic dynamic systems in a  long enough interval.

On the other side, although Lorenz's  famous claim that ``long-term prediction of chaos is impossible''  has been widely accepted by scientific community, the word ``long-term'' is not very clear. Is one day long enough? Or millions of years?  Obviously,  a time-scale is needed for us to understand Lorenz's claim better.

Currently, a safe numerical approach, namely the ``Clean Numerical Simulation'' (CNS) \cite{Liao2012, Liao2013}, is proposed to gain  mathematically reliable computer-generated  trajectories of chaotic dynamic system in a finite but long enough interval.    The CNS is based on Taylor series method (TSM) \cite {Corliss1982, Barrio2005} at high enough order of approximation and the high precision data with large enough number of significant digits.   The Taylor series method \cite{Corliss1982, Barrio2005} is one of the oldest method, which can trace back to Newton, Euler, Liouville and Cauchy.  It has an advantage that its formula at arbitrarily high order can be easily expressed in the same form.  So, from viewpoint of numerical simulations,  it is rather easy to use the Taylor series method at very high order so as to deduce the truncation error to a required level.  Besides, the round-off error can be reduced to arbitrary level by means of computer algebra system (such as Mathematica) or the multiple precision (MP) library \cite{MP}.   Thus, using the CNS,  the numerical noises can be decreased to such a small level that both truncation and round-off errors are negligible in a given finite but long enough interval.   For example, using the CNS  ($\Delta t=0.01$) with the 400th-order Taylor series method and 800-digit precision data, Liao \cite{Liao2009} gained, for the first time, the mathematically reliable chaotic simulation of Lorenz equation in the interval $0 \leq t \leq 1000$  LTU  (Lorenz time unit) by means of Mathematica.  In 2011, Wang et al. \cite{Wang2011} greatly  decreased  the  required  CPU  times of the CNS by employing a parallel algorithm and the multiple precision (MP) library of C.     Using the parallel CNS with the 1000th-order Taylor expansion and the 2100 digit multiple precision,  Wang et al.  \cite{Wang2011}  gain a mathematically reliable chaotic simulation of Lorenz equation in the interval [0,2500]  within only 30 hours, which is validated using a more accurate simulation given by the 1200th-order Taylor expansion and the 2100 digit multiple precision.    Their results  \cite{Wang2011}  confirm the correction and reliability of Liao's simulation \cite{Liao2009} in the interval $[0,1000]$.  Currently,  using  1200 CPUs  of the National Supercomputer TH-A1 and  the  modified  parallel integral algorithm based on  the CNS with the  3500th-order Taylor expansion  and  the 4180-digit  multiple precision data,  Liao and Wang  \cite{Liao-Wang} gain a mathematically reliable simulation of  chaotic solution of Lorenz equation in a rather long interval $[0,10000]$.  All of these illustrate that,  the uncertainty of simulations of chaotic trajectories (in a given, finite but long enough interval)  caused  by  numerical  noises  can  be  avoided by means of the CNS.   Thus, from {\em mathematical} viewpoint,  given an {\em exact} initial condition,  we  can  gain  {\em mathematically}  reliable  trajectories  of  chaotic dynamic systems in a {\em finite} but long {\em enough} interval by means of the CNS, without any observable uncertainty of simulation.

Is such a {\em mathematically} reliable chaotic trajectory (in a finite but long enough interval) {\em physically}  correct?

 Note that the uncertainty of simulations of chaotic trajectories is caused by many factors.  Theoretically speaking, given an {\em exact} initial condition, the uncertainty is completely caused by numerical noises, i.e. truncation and round-off error, where truncation error is determined by numerical algorithms and round-off error is due to the limited precision of numerical data, respectively.  However, in practice,  initial conditions are {\em not} exact in practice: they contain both artificial and physical uncertainty.  The artificial uncertainty mainly comes from limited precision of measurement.  The physical uncertainty is due to the inherently uncertain/random  property of nature,  caused by such as thermal fluctuation,   wave-particle duality of de Broglie's wave, and so on.   Generally,  the artificial uncertainty is much larger than the physical uncertainty.   So,  for dynamic systems,  physical uncertainty determines a time of physical limit of  prediction,  denoted by $T_p^{max}$,  beyond which trajectories are essentially uncertain in {\em physics}, as illustrated in this paper.   In order to investigate the  time of physical limit of prediction, we {\em assume} that there is no artificial uncertainty,  i.e.   measurement   can have  precision of arbitrary degree.  In this way, the uncertainty of initial conditions cased by limited precision of measurement is avoided, too.  Thus, we  can  focus on the case  that  initial conditions contain  micro-level  physical  uncertainty  only,  which are often in the (dimensionless) micro-level of $10^{-20}$ to $10^{-60}$ or even smaller.  It should be emphasized that the uncertainty of initial condition is in the micro-level level, but the considered dynamic system is about macroscopic phenomena.  Thus, this is a problem with multiple scales.  Fortunately,  the propagation of such kind of micro-level uncertainty can be reliably and accurately simulated by means of the CNS now, as illustrated by Liao \cite{Liao2013}.

Without loss of generality, we consider here the famous three-body problem \cite{Poincare1890, Henon1964, Diacu1996, Valtonen2005} with chaotic motion, governed by Newtonian gravitational  law.   In Section 2 we briefly describe the numerical algorithms based on the CNS approach.    According to the string theory \cite{Polchinshi1998}, the Planck length is the order of magnitude of oscillating strings that form elementary particles, and shorter length do not make physical senses.  Thus,  as pointed out in Section 3, the physical uncertainty of initial position of the three bodies might be in the (dimensionless) level of $10^{-60}$.   The propagation of such kind of micro-level uncertainty for the three-body problem in a finite but long enough interval, together with the mathematically reliable chaotic trajectories,  can be  accurately simulated by means of the CNS, as illustrated by Liao \cite{Liao2013}.     It should be emphasized that such a micro-level uncertainty of initial position is inherent and unavoidable in physics, i.e. objective.     Our reliable simulations given in Section 4  suggest that such tiny physical uncertainty might lead to the macroscopic uncertainty of chaotic trajectories beyond a limit time $T^{max}_p$, where $T_p^{max}\approx 810$, although all of these chaotic trajectories are mathematically reliable in the interval [0,1000].  In other word, considering the physical  uncertainty of initial position at  the micro-level $10^{-60}$,   we  can  gain chaotic  trajectories in the interval  $[0, T^{max}_p]$ without obvious difference,  but beyond it  these mathematically reliable  chaotic trajectories have obvious difference  and  thus the  system  becomes  essentially  uncertain  in  physics.   This is mainly because the micro-level physical uncertainty of initial condition transfers into the macroscopic uncertainty when $ t > T_p^{max}$, so that it is impossible to make a {\em physically} correct prediction beyond it.   Thus, $T_p^{max}\approx 810$ is the time of the  {\em physical limit of prediction} of  chaotic trajectories  for the special case of three-body problem considered in this paper, which provides us a time-scale for at most how long mathematically reliable  simulations  of  chaotic motion of the three bodies are physically correct.   Finally, concluding remarks and some discussions about the concept ``physical limit of prediction'' are given in Section 5.

\section{Numerical algorithms of the CNS}

Applications of the CNS to the Hamiltonian H\'{e}non-Heiles system for motion of stars orbiting in a plane  suggest that there exist a physical limit of prediction time, beyond which trajectories of chaotic dynamic systems become essentially uncertain in physics \cite{Liao2013}.
Note that  the Hamiltonian H\'{e}non-Heiles system is a simplified model for motion of stars orbiting in a plane.    However,  orbits of stars are three dimensional in practice.    Thus, it is necessary to investigate some more accurate physical models, such as the famous three-body problem \cite{Poincare1890, Henon1964, Diacu1996, Valtonen2005}  governed by the Newtonian gravitation law.   In fact,  non-periodic  results  were  first  found  by Poincar\'{e} \cite{Poincare1890} for three-body problem.

Let us consider the famous  three-body problem, say, the motion of three celestial bodies under their mutual gravitational attraction.    Let $x_1, x_2, x_3$ denote the three orthogonal axises.  The  position  vector  of  the $i$ body is expressed by
${\bf r}_i = (x_{1,i},x_{2,i},x_{3,i})$.   Let $T$ and $L$ denote the characteristic time and length  scales,   and $m_i$  the mass of the $i$th  body,  respectively.    Using Newtonian gravitation law,  the motion of the three bodies are governed by the corresponding non-dimensional equations
\begin{equation}
\ddot{x}_{k,i} = \sum_{j=1,  j\neq i}^{3}  \rho_j \frac{(x_{k,j}-x_{k,i})}{R_{i,j}^{3}},\;\;\; k = 1,2,3,  \label{geq:x[k,j]}
\end{equation}
where
\begin{equation}
R_{i,j} = \left[ \sum_{k=1}^{3} (x_{k,j}-x_{k,i})^2 \right]^{1/2}   \label{def:R[i,j]}
\end{equation}
and
\begin{equation}
 \rho_i = \frac{m_i}{m_1}, \;\; i = 1,2,3
\end{equation}
denotes the ratio of the mass.

In the frame of the CNS,  we use  the $M$-order Taylor series method
\begin{equation}
x_{k,i}(t) \approx \sum_{m=0}^{M} \alpha_m^{k,i} \; (t-t_0)^m \label{series:x[k,i]}
\end{equation}
to accurately calculate the orbits of the three bodies, where the coefficient  $\alpha_m^{k,i}$ is only dependent upon the time $t_0$.     Note that the position  $x_{k,i}(t)$ and velocity $\dot{x}_{k,i}(t)$ at $t=t_0$ are known, i.e.
\begin{equation}
\alpha_0^{k,i}  = x_{k,i}(t_0),  \;\;  \alpha_1^{k,i}  = \dot{x}_{k,i}(t_0).
\end{equation}
 The recursion formula of $\alpha_m^{k,i}$ for $m\geq 2$ is derived from (\ref{geq:x[k,j]}), as described  below.

To apply the high-order Taylor series method efficiently, we should give explicit formulas for the coefficient $\alpha_m^{k,i}$.   Write  $1/R_{i,j}^3$ in the form
\begin{equation}
f_{i,j} = \frac{1}{R_{i,j}^3} \approx \sum_{m=0}^{M} \beta_m^{i,j} \; (t-t_0)^m  \label{series:1/R[i,j]}
\end{equation}
with the symmetry property $\beta_m^{i,j} =\beta_m^{j,i}$, where $\beta_m^{i,j} $ is determined later.
Substituting (\ref{series:x[k,i]}) and (\ref{series:1/R[i,j]}) into (\ref{geq:x[k,j]}) and comparing the like-power of $(t-t_0)$, we have the recursion  formula
\begin{equation}
\alpha_{m+2}^{k,i} = \frac{1}{(m+1)(m+2)} \sum_{j=1, j\neq i}^{3} \rho_j \sum_{n=0}^{m} \left( \alpha_n^{k,j}-\alpha_n^{k,i}\right) \; \beta_{m-n}^{i,j}, \;\; m \geq 0.
\end{equation}
Thus, the positions and velocities of the three bodies  at the next time-step $t_0+\Delta t$ read
\begin{eqnarray}
x_{k,i}(t_0+\Delta t) &\approx & \sum_{m=0}^{M} \alpha_m^{k,i} \; (\Delta t)^m  \label{result:position} ,\\
\dot{x}_{k,i}(t_0+\Delta t) &\approx & \sum_{m=1}^{M} m \; \alpha_m^{k,i} \; (\Delta t)^{m-1} \label{result:x[k,i]} . \label{result:velocity}
\end{eqnarray}

Write
\begin{equation}
 S_{i,j} = R^6_{i,j} \approx \sum_{m=0}^{M} \gamma_m^{i,j} \; (t-t_0)^m, \;\;  f_{i,j}^2 = \sum_{m=0}^{M} \sigma_{m}^{i,j} (t-t_0)^m,  \label{series:S[i,j]}
 \end{equation}
 with the symmetry property $\gamma_m^{i,j}=\gamma_m^{j,i}$ and $\sigma_m^{i,j}=\sigma^{j,i}_m$.   Substituting (\ref{def:R[i,j]}), (\ref{series:x[k,i]}) and (\ref{series:1/R[i,j]}) into the above definitions and comparing the like-power of $(t-t_0)$,  we have
\begin{eqnarray}
\gamma^{i,j}_m &=& \sum_{n=0}^{m} \mu^{i,j}_{m-n} \sum_{k=0}^{n}\mu_k^{i,j} \mu_{n-k}^{i,j},    \\
\sigma_m^{i,j} &=& \sum_{n=0}^m \beta^{i,j}_n \; \beta^{i,j}_{m-n},
\end{eqnarray}
with
\begin{equation}
\mu^{i,j}_m = \sum_{k=1}^{3} \sum_{n=0}^{m}\left( \alpha^{k,j}_{n}-\alpha^{k,i}_{n}\right)\left( \alpha^{k,j}_{m-n}-\alpha^{k,i}_{m-n}\right), \;\; i\neq j, \; m\geq 1,
\end{equation}
and the symmetry   $\mu_m^{i,j}=\mu_m^{j,i}$.
Using the definition (\ref{series:1/R[i,j]}),  we have
\[   S_{i,j} f_{i,j}^2 = 1.   \]
Substituting  (\ref{series:S[i,j]}) into the above equation  and comparing the like-power of $(t-t_0)$, we have
\[    \sum_{n=0}^{m} \gamma_n^{i,j} \; \sigma_{m-n}^{i,j} = 0,  \;\; m\geq 1, \]
which gives the recursion formula
\begin{equation}
\beta_m^{i,j} = -\frac{1}{2\beta_0^{i,j} \gamma_0^{i,j}} \left\{\sum_{n=1}^{m} \gamma_n^{i,j}\sigma_{m-n}^{i,j} +\gamma_0^{i,j}\sum_{k=1}^{m-1}\beta_k^{i,j}\beta_{m-k}^{i,j}\right\}, \;\; m\geq 1, \; i\neq j.
\end{equation}
 In addition,  it is straightforward that
\begin{equation}
\beta_0^{i,j} = \frac{1}{R_{i,j}^3}, \;\;\;  \mu_0^{i,j} = R_{i,j}^2, \;\;\; \sigma_0^{i,j} = \left( \beta_0^{i,j} \right)^2,  \;\;\;  \gamma^{i,j}_0 =  \left( \mu_0^{i,j}\right)^3,\;\;\; \mbox{at $t=t_0$}.
\end{equation}

It is a common knowledge that numerical methods always contain truncation  and round-off errors.    To decrease the round-off error,  we express  the positions,  velocities, physical parameters and {\em all} numerical data  in $N$-digit precision\footnote{Mathematica is used in this article},  where $N$ is a large enough positive integer.   Obviously, the larger the value of $N$, the smaller the round-off error.    Besides, for a reasonable time step $\Delta t=t-t_0$,
 the higher the order $M$ of Taylor expansion (\ref{series:x[k,i]}),  the smaller the truncation error.    Therefore,  if the order $M$ of Taylor expansion (\ref{series:x[k,i]}) is high enough and all data are expressed in  accuracy of long enough digits,   both of truncation and round-off errors of the above-mentioned CNS approach (with reasonable time step $\Delta t$) can be so small that numerical noises are negligible in a (given) finite but long enough time interval, say,  we can gain mathematically reliable chaotic simulations in a given time interval.    In this way,  the mathematically reliable chaotic orbits of the three bodies can be gained in a given, finite but long enough interval.

In this article, the computer algebra system Mathematica is employed.   By means of the Mathematica,  it is rather convenient to express all datas in 300-digit  precision, i.e. $N=300$.   In this way, the round-off error is so small that it is almost negligible in the given time interval ($0\leq t\leq 1000$).  And the accuracy of the CNS  simulations   increases as the order $M$  of Taylor expansion (\ref{series:x[k,i]}) enlarges, as shown in the next section.

In summary,  the  uncertainty of  chaotic trajectories  caused  by  the  numerical noises  can be avoided by means of the CNS, so that the propagation of the micro-level physical uncertainty of initial condition can be investigated accurately.

\section{Physical uncertainty of initial conditions}

It is widely accepted  that  microscopic  phenomena are essentially uncertain/random.    Let us  first consider some typical length scales of microscopic  phenomena which are widely used in modern physics.     For example,
Bohr radius
\[ r = \frac{\hbar^2}{m_e \; e^2} \approx 5.2917720859(36) \times 10^{-11} \;\; \mbox{(m)}\]
 is the approximate size of a hydrogen atom, where $\hbar$ is a reduced Planck's constant, $m_e$ is the electron mass, and $e$ is the elementary charge, respectively.   Besides,  the so-called  Planck length
\begin{equation}
l_P = \sqrt{\frac{\hbar \; G}{c^3}} \approx 1.616252(81) \times 10^{-35} \;\;\; \mbox{(m)}
\end{equation}
is the length scale at which quantum mechanics, gravity and relativity all interact very strongly,  where  $c$ is the speed of light in a vacuum, and $G$ is the gravitational constant, respectively.   According to the string theory \cite{Polchinshi1998}, the Planck length is the order of magnitude of oscillating strings that form elementary particles,  and  {\em shorter length do not make physical senses}.    Especially,  in some forms of quantum gravity, it becomes  {\em impossible}  to  {\em determine} the difference between two locations less than one Planck length apart.    Therefore,  in the  level  of the Planck length,  position of a body is  {\em inherently} uncertain.    This kind of microscopic  physical  uncertainty is {\em inherent} and has nothing to do with  Heisenberg uncertainty principle \cite{Heisenberg1927}, say, it is {\em objective}.

In addition,  according to de Broglie \cite{Broglie1924},  any  a  body has the so-called  wave-particle duality.   The de Broglie's wave of a body has non-zero amplitude.  Thus,  position of a body is uncertain:  it could be almost {\em anywhere}   along de Broglie's wave packet.   Thus,  according to  the de Broglie's wave-particle duality,  position of a star/planet is {\em inherent} uncertain, too.   Therefore,  it is  reasonable   to  assume that, from the physical viewpoint,   the micro-level inherent fluctuation of position of a body  shorter than the  Planck length $l_p$ is essentially  uncertain and/or  random.

To make the  Planck length $l_p\approx 1.62 \times 10^{-35}$ (m)  dimensionless, we use the diameter of Milky Way Galaxy as the characteristic length,  say,  $d_M  \approx 10^5$ light year $\approx 9 \times 10^{20}$ meter.  Obviously,  $l_p/d_M \approx 1.8 \times 10^{-56} $ is a rather small dimensionless number.   Thus, as mentioned above,  two (dimensionless) positions shorter than $10^{-56}$ do not make physical senses in physics.  So,  it is reasonable to assume  that  the inherent  uncertainty of the dimensionless position  of a star/planet  is  in  the micro-level $10^{-60}$.   Therefore, the tiny difference $d {\bf r}_1 = \pm 10^{-60} (1,0,0)$ of the initial conditions is in the micro-level:  the difference is so small that all of these initial conditions can be regarded as the {\em same} in physics!

Mathematically,   $10^{-60}$ is a tiny number,  which  is  much  smaller  than truncation and round-off errors of {\em traditional} numerical  algorithms  based on 16 or 32-digit precision.    So, it is impossible to investigate the influence and evaluation of this inherent micro-level uncertainty of initial conditions by means of these traditional numerical methods.  However,  the micro-level uncertainty $10^{-60}$ is much larger than the truncation and round-off errors of the CNS  simulations by means of  the high-order Taylor expansion and data in  300-digit precision with a reasonable time step $\Delta t$, as illustrated below.   This is mainly because  the CNS provides us a safe  tool to study the propagation of such kind of inherent micro-level uncertainty of initial conditions, as currently illustrated in \cite{Liao2013}.

\section{Reliable long-term simulations of chaotic motion}

 As mentioned before, the initial conditions contain both of artificial and physical uncertainty.   For the sake of simplicity,  the artificial uncertainty due to limited precision of measurement  is assumed to be zero.  In this way, we can focus on the propagation of the micro-level physical uncertainty of initial conditions.

Without loss of generality,  let us consider the motion of three bodies  with the initial positions
\begin{equation}
 {\bf r}_1 = (0 ,0,-1) + d{\bf r}_1, \;   {\bf r}_2 = (0,0,0),  {\bf r}_3 = - ({\bf r}_1  + {\bf r}_2),  \label{ic:r}
\end{equation}
and the initial velocities
\begin{equation}
\dot{\bf r}_1 = (0,-1,0), \dot{\bf r}_2 = (1,1,0), \dot{\bf r}_3 = -(\dot{\bf r}_1 + \dot{\bf r}_2),  \label{ic:velocity}
\end{equation}
where $ d {\bf r}_1 = \delta (1,0,0) $ is the term with the micro-level physical uncertainty.   For simplicity,  we first only focus on the following  three cases:  $\delta  = 0$,   $\delta =  +10^{-60}$ and $\delta =  - 10^{-60}$.     {\em Mathematically},  the  three initial positions have the tiny difference in the level of $10^{-60}$ (i.e. the physical uncertainty).  However,  the difference is even smaller than the Planck length so that all of these initial conditions can be regarded as the {\em same} in {\em physics}, since any lengths shorter than the Planck length do not make physical senses  \cite{Polchinshi1998}.   For the sake of simplicity,  let us consider the case of  equal masses, i.e. $\rho_j = 1$ ($j=1,2,3$).     We are interested in the chaotic orbits of the three bodies in the time interval $0\leq t \leq 1000$, which is long enough for our investigation interest, as shown below.

Note that the initial conditions satisfy
\[    \sum_{j=1}^3 \dot{\bf r}_j(0)  = \sum_{j=1}^3 {\bf r}_j(0) = 0.  \]
Thus, due to the momentum conversation, we have
\begin{equation}
\sum_{j=1}^3 \dot{\bf r}_j(t)  = \sum_{j=1}^3 {\bf r}_j(t) = 0, \hspace{1.0cm} t \geq 0  \label{conversation:r}
\end{equation}
in general.

All data are expressed in 300-digit precision, i.e. $N=300$, using Mathematica.  Thus, the round-off error is  very small  (compared to the physical uncertainty at the level $10^{-60}$)  and thus negligible.
In addition, the higher the order $M$ of Taylor expansion (\ref{series:x[k,i]}),  the smaller the truncation error, i.e.  the more accurate the results at $t=1000$.     Assume that,  at $t=1000$,  we have   the result $x_{1,1} = {\bf 1.81510} 12345$    by means of the $M_1$-order Taylor expansion and the result $x_{1,1} = {\bf 1.81510} 47535$  by means of the $M_2$-order Taylor expansion,  respectively,  where $M_2 > M_1$.  Then,  the result $x_{1,1}$  by means of the lower-order ($M_1$) Taylor expansion  is said to be in the accuracy of 5 significance digit,  expressed by $n_s = 5$.       For more details about the CNS, please refer to Liao \cite{Liao2013}.

\begin{figure}
\centering
\includegraphics[scale=0.4]{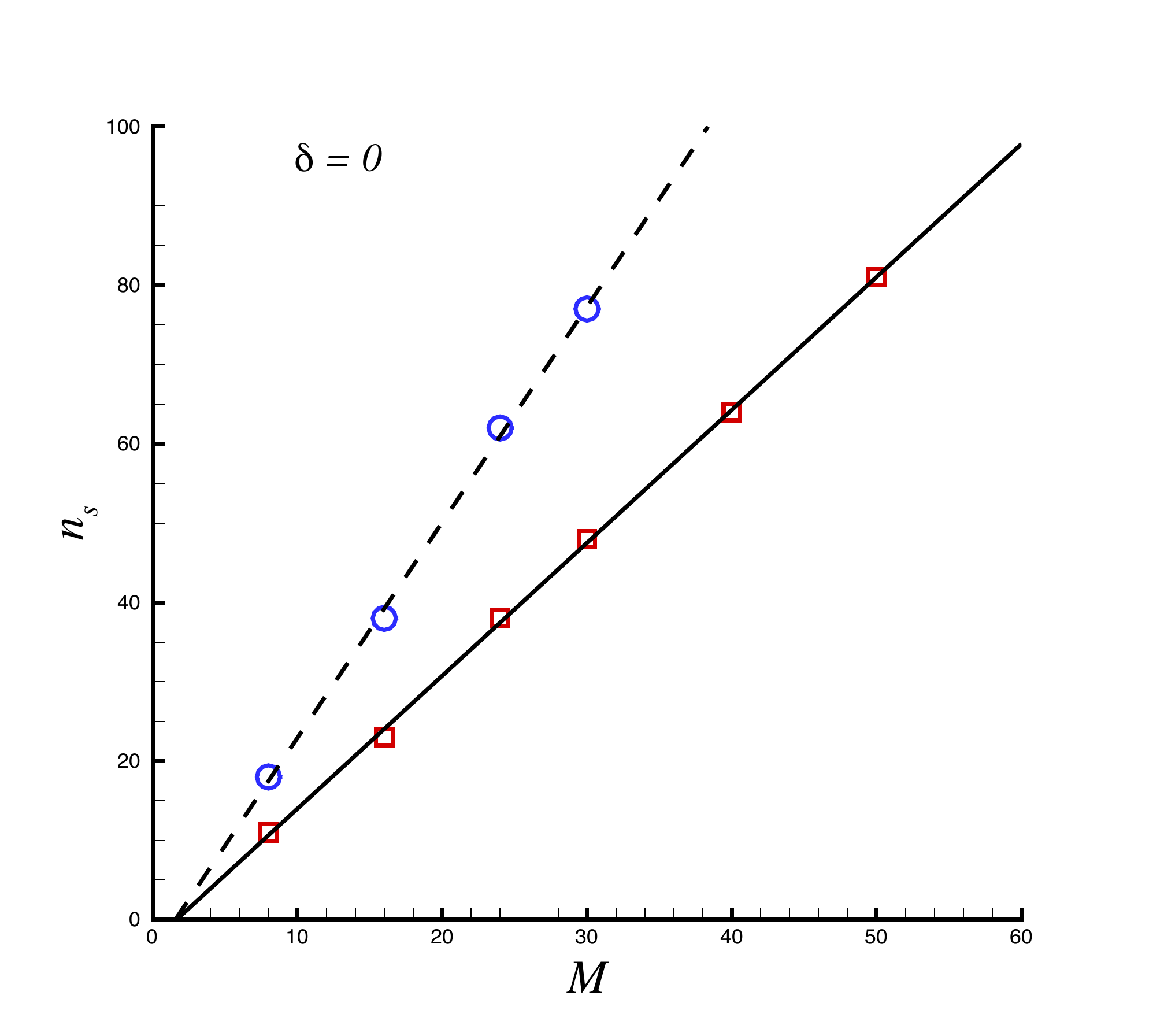}
\caption{ Accuracy (expressed by $n_s$, number of significance digits) of the CNS results at $t=1000$  in the case of $\delta = 0$ by means of $N=300$, the different time-step $\Delta t$ and the different order ($M$) of Taylor expansion.  Square: $\Delta t = 10^{-2}$; Circle: $\Delta t = 10^{-3}$.  Solid line: $n_s \approx 1.6762M - 2.7662$; dashed line: $n_s \approx  2.7182 M - 4.2546$.   }
\label{figure:accuracy:delta-0}
\end{figure}

\begin{figure}
\centering
\includegraphics[scale=0.3]{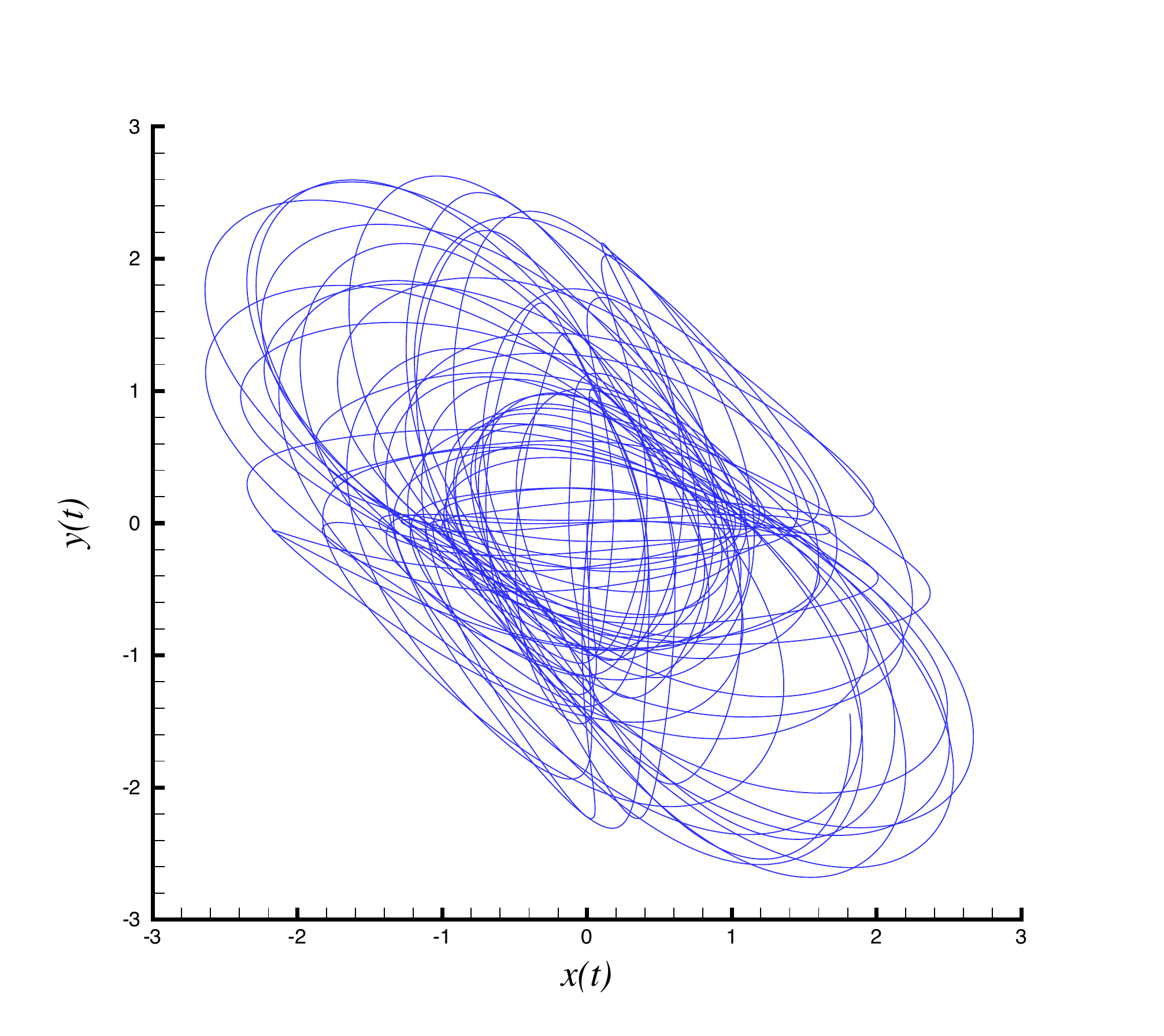}
\includegraphics[scale=0.3]{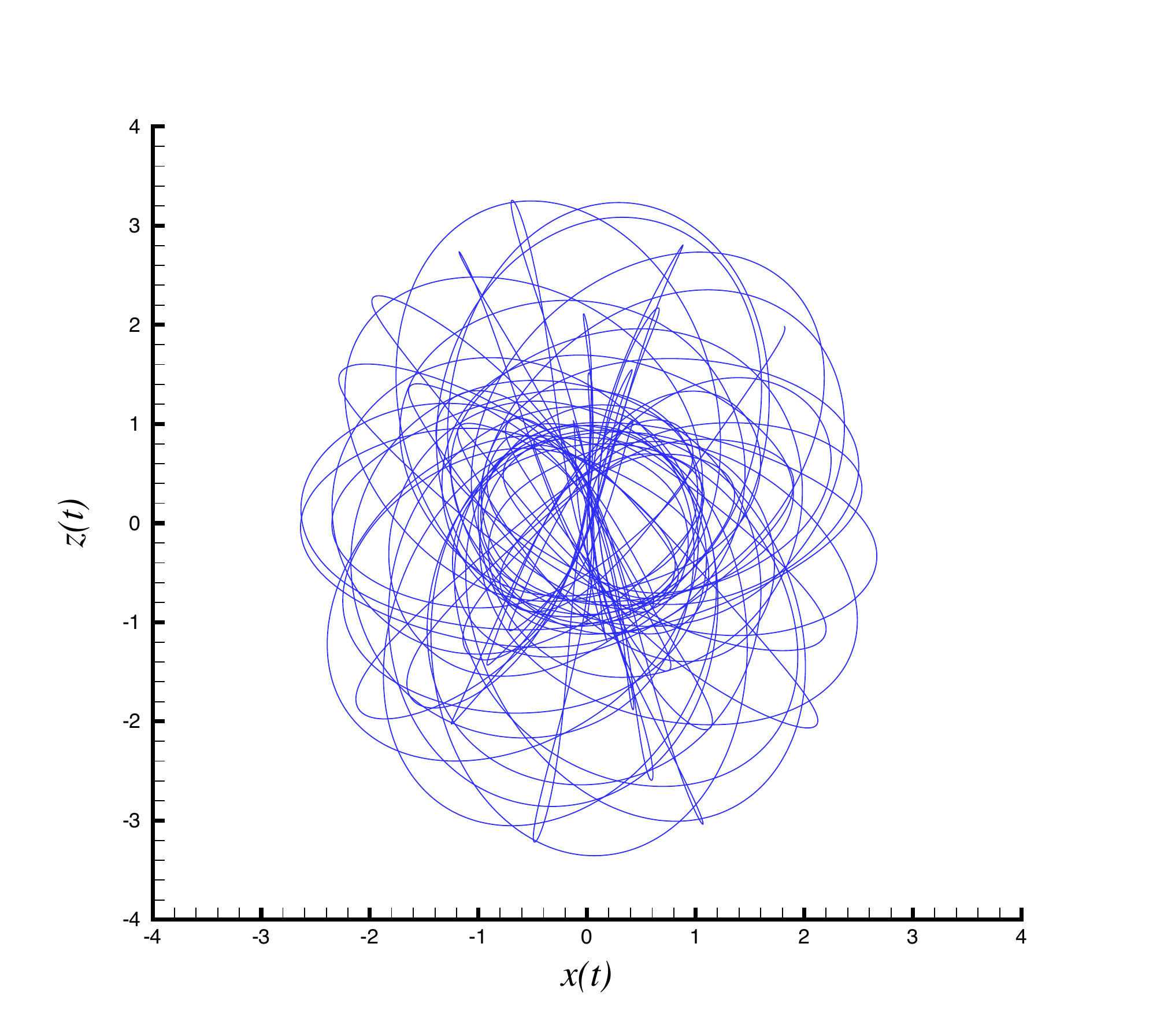}
\caption{ $x-y$ and $x-z$ of Body 1  ($0 \leq t \leq 1000$)  in the case of $\delta = 0$.}
\label{figure:body1-2D}
%\end{figure}

%\begin{figure}
\centering
\includegraphics[scale=0.3]{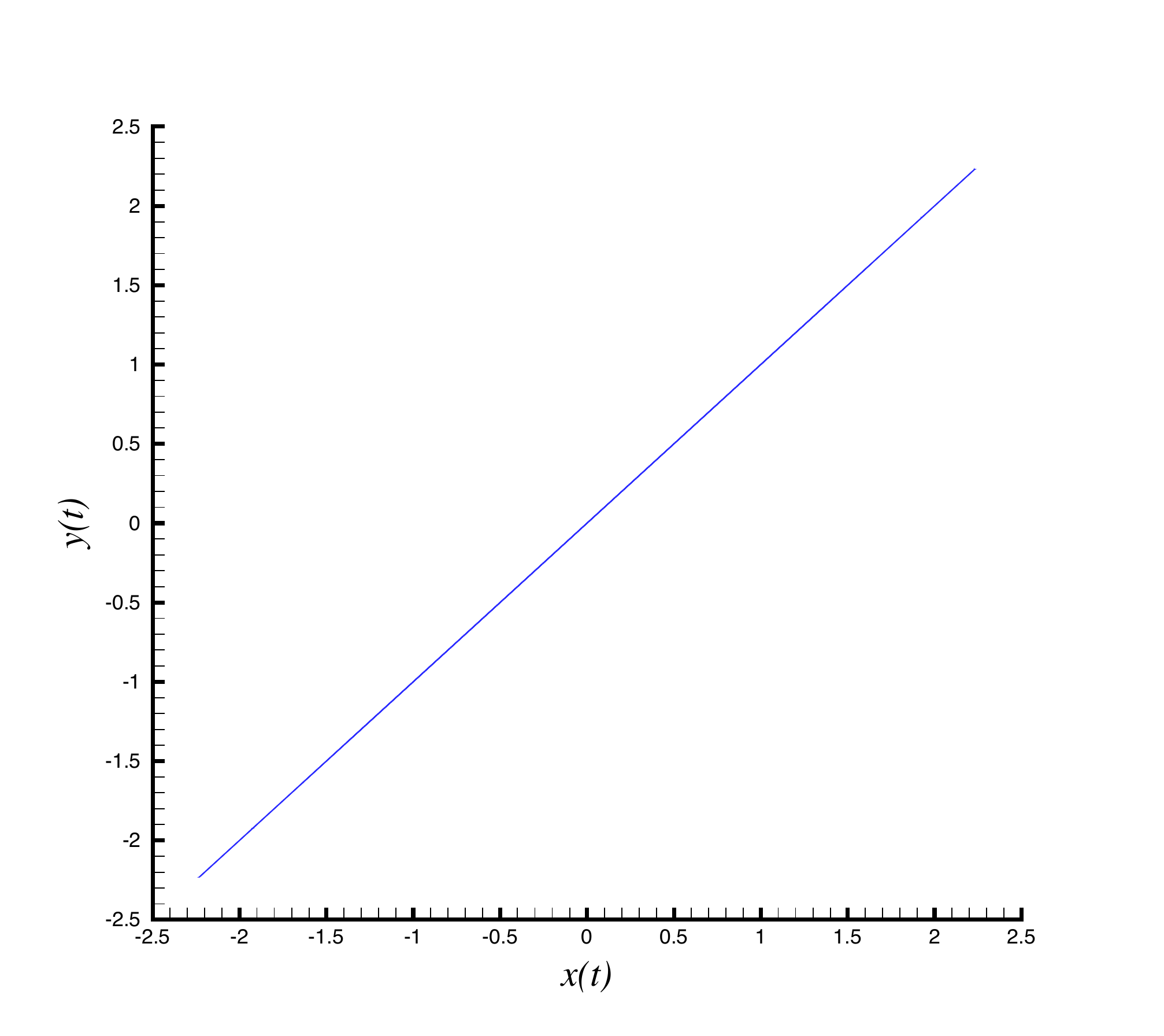}
\includegraphics[scale=0.3]{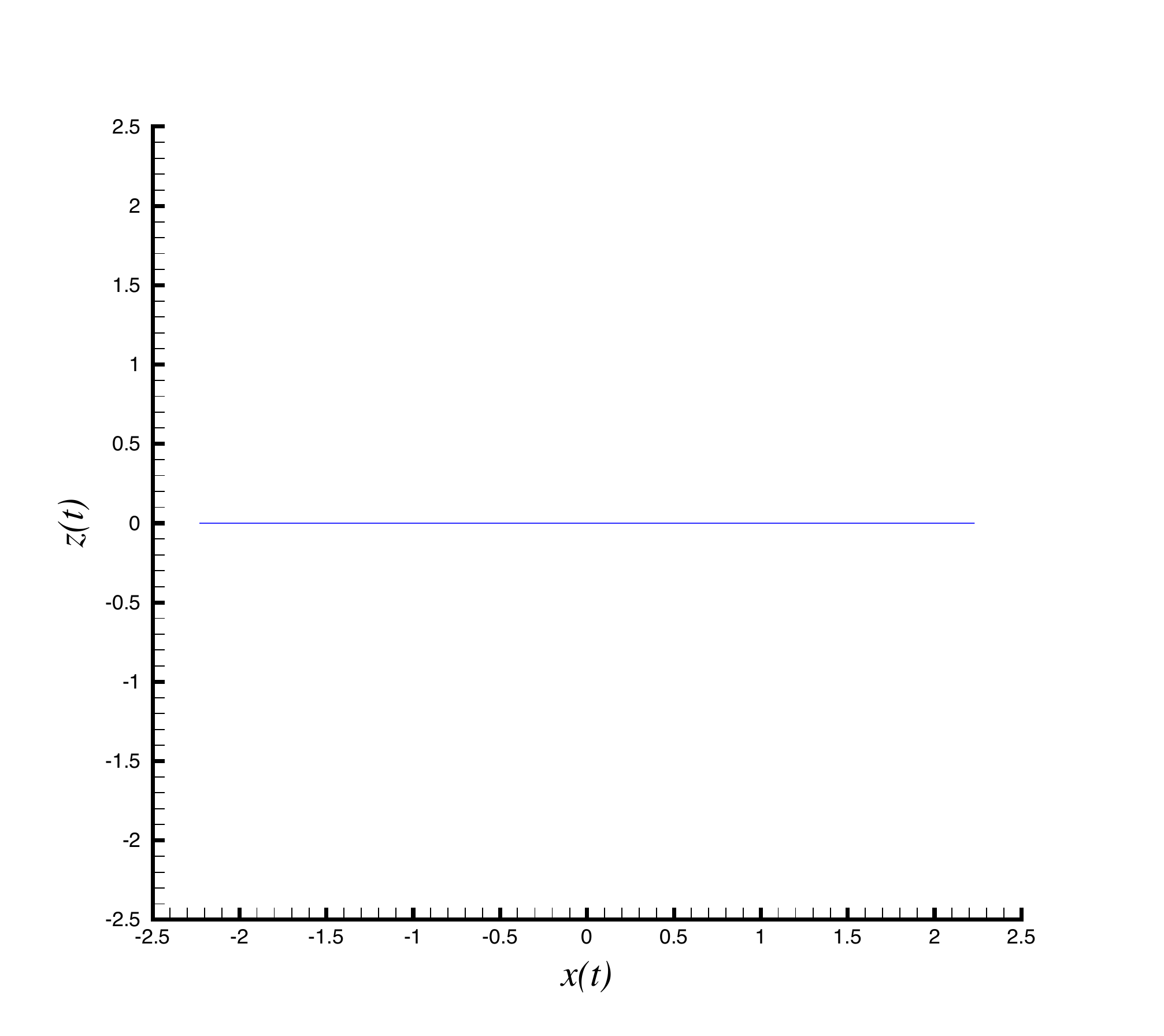}
\caption{ $x-y$ and $x-z$ of Body 2  ($0 \leq t \leq 1000$) in the case of $\delta = 0$. }
\label{figure:body2-2D}
%\end{figure}

%\begin{figure}
\centering
\includegraphics[scale=0.3]{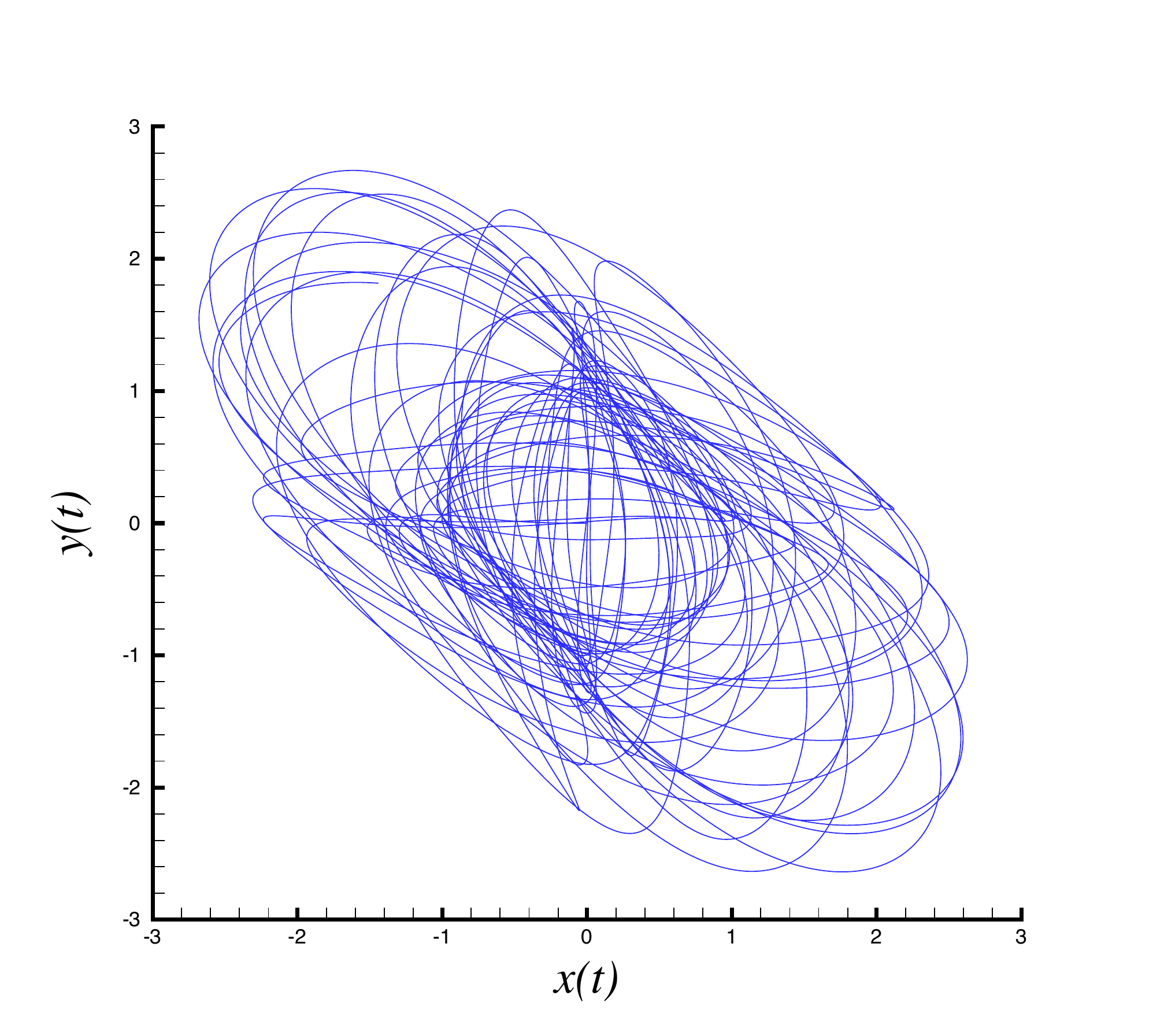}
\includegraphics[scale=0.3]{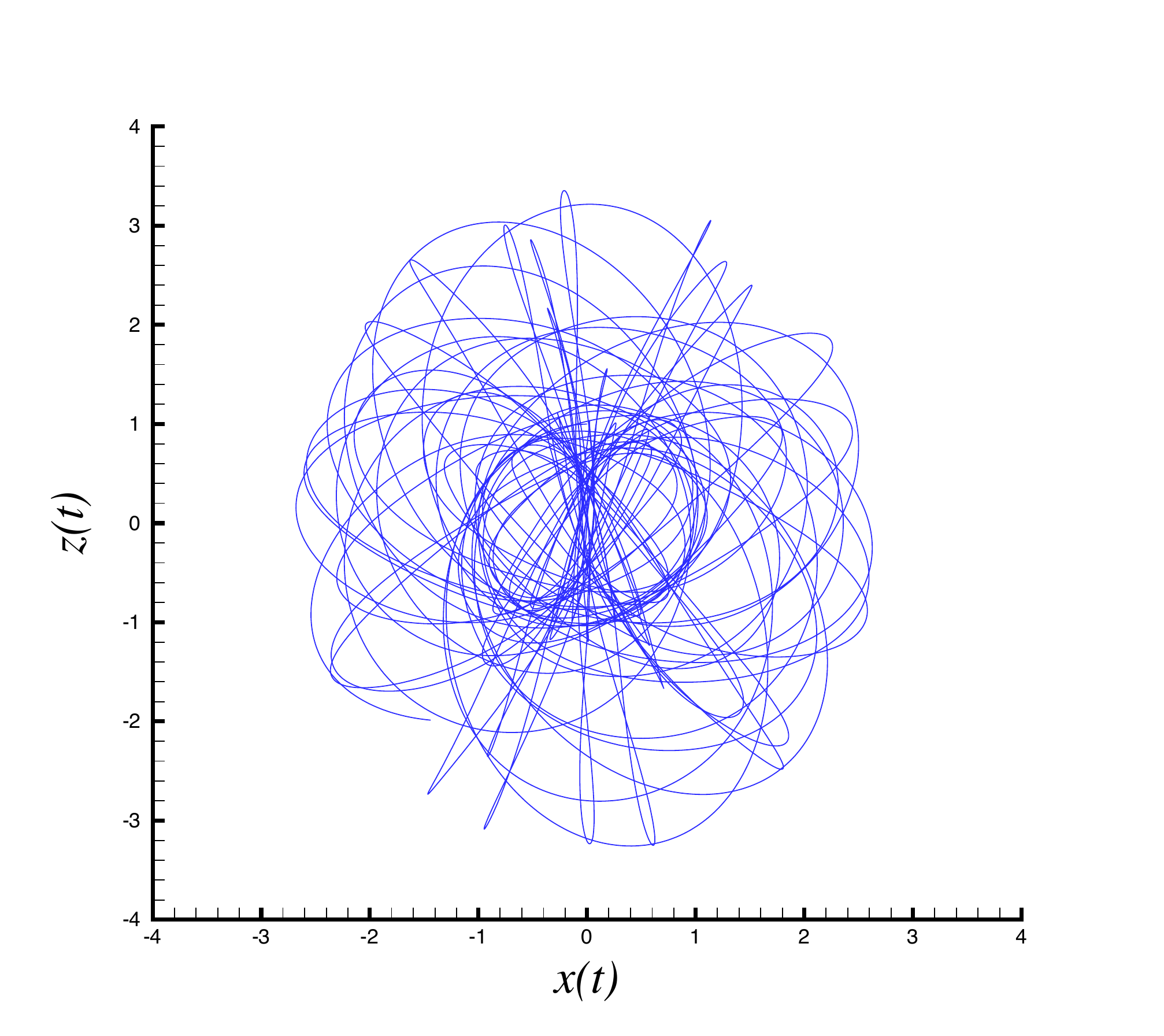}
\caption{ $x-y$ and $x-z$ of Body 3  ($0 \leq t \leq 1000$)  in the case of $\delta = 0$. }
\label{figure:body3-2D}
\end{figure}

When $\delta = 0$, the corresponding three-body problem has chaotic orbits with the Lyapunov exponent  $\lambda = 0.1681$, as pointed by Sprott \cite{Sprott2010} (see  Figure 6.15 on page 137).    It is well-known that a chaotic dynamic system has  the sensitivity dependence on initial condition ( SDIC).    Thus, in order to gain reliable   numerical simulations of the  chaotic orbits  in such a long interval $0\leq t \leq 1000$,  we employ  the  CNS approach using the high-enough order $M$  of Taylor expansion with all data  expressed in 300-digit precision.

It is found that, when $\delta=0$,  the CNS  simulations using $\Delta t = 10^{-2}$, $N = 300$ and $M= 8,16, 24, 30, 40$ and 50   agree  each other   in the {\em whole}  interval [0,1000]  at least in the accuracy of 11,  23,  38,  48,  64  and 81  significance digits, respectively.    Approximately,  $n_s$, the number of significance digits of the positions at $t=1000$,   is linearly  proportional to $M$ (the order of Taylor expansion), say, \[ n_s \approx 1.6762 M - 2.7662,\]  as shown in Fig.~\ref{figure:accuracy:delta-0}.     For example,  the CNS simulation using the 50th-order Taylor expansion and data in 300-digit precision (with $\Delta t=10^{-2}$)  provides us  the position of Body 1  at  $t=1000$   in the accuracy of 81 significance digit:
\begin{eqnarray}
x_{1,1} &=&
  +1.81510 47535 62951 61721 65940 08845 44006 45690 \nonumber \\ && 03032 05574  32375 90103
28524 04361 81354 98683 4,  \label{x[1,1]-delta-0}\\
x_{2,1} &= & -1.44063 51440 58286 16113 38350 55406  74890 01231 \nonumber \\ && 29002 84853
72197 68908 17632 27032 47482 40993 8, \\
x_{3,1} & = & + 1.98700 78875 78629 88109 76776 41498 57789  79168 \nonumber \\ &&  46299 74639 70517 57074.
11773 08217 34512 84475 9. \label{x[3,1]-delta-0}
\end{eqnarray}
Using the smaller time step $\Delta = 10^{-3}$ and data in 300-digit precision (i.e. $N=300$),  the CNS simulations given by the 8, 16, 24 and 30th-order Taylor expansion agree in the whole interval [0,1000] in the accuracy of 18, 38, 62 and 77 significance digits, respectively.  Approximately,  $n_s$, the number of significance digits of the positions at $t=1000$,   is linearly  proportional to $M$ (the order of Taylor expansion), say, \[ n_s \approx 2.7182 M - 4.2546,\]  as shown in Fig.~\ref{figure:accuracy:delta-0}.   It should be emphasized that the CNS simulations given  by $\Delta t = 10^{-3}$ and $M=30$ agree (at least)  in the 77 significance digits  with those by $\Delta t = 10^{-2}$ and  $M=50$ in the {\em whole} interval $[0,1000]$.   In addition, the momentum conservation (\ref{conversation:r}) is satisfied in the level of $10^{-295}$.   All of these  confirm the mathematical correction and reliability  of our  CNS simulations\footnote{Liao \cite{Liao2013} proved a convergence-theorem and explained the validity and reliability of the CNS by using the mapping $x_{n+1} = \mbox{mod}(2 x_n,1)$.}.   Thus, although the  considered  three-body problem has chaotic orbits,   we are quite sure that our numerical simulations given by the CNS  using  the 50th-order Taylor expansion and accurate data in 300-digit precision (with $\Delta t=10^{-2}$) are {\em mathematically} reliable in the accuracy of 77 significance digits  in the {\em whole} interval $0\leq t \leq 1000$.

The orbits of the three bodies in the case of $\delta=0$ are as shown in Figs.~\ref{figure:body1-2D} to \ref{figure:body3-2D}.    The orbits of Body~1 and Body~3 are chaotic.  This agrees well with Sprott's conclusion  \cite{Sprott2010} (see  Figure 6.15 on page 137).   However, it is interesting that Body~2  oscillates along a line on the plane $z = 0$.    So, since  $\sum_{j=1}^3 \dot{\bf r}_j$   = $\sum_{j=1}^3{\bf r}_j$   = 0 due to the momentum conservation,  the  chaotic  orbits  of  Body~1 and Body~3 must be  symmetric  about the  regular  orbit  of Body~2.   Thus, although the orbits of Body~1 and Body~3  are  disorderly, the three bodies as a system have an elegant symmetry.

 \begin{figure}
\centering
\includegraphics[scale=0.4]{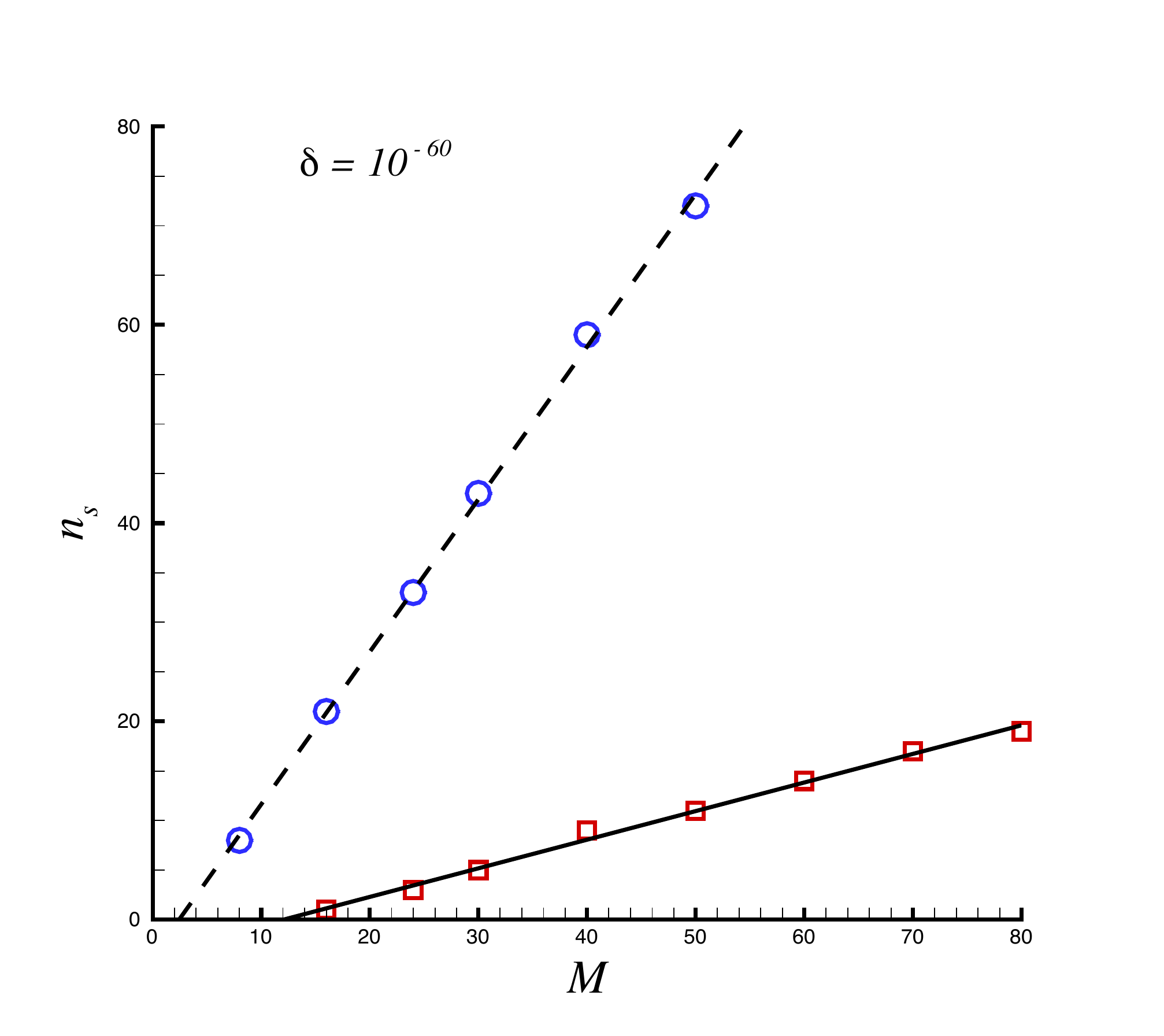}
\caption{ Accuracy (expressed by $n_s$, number of significance digits) of the result at $t=1000$  in the case of $\delta = 10^{-60}$ by means of $N=300$, the different time-step $\Delta t$ and the different order of Taylor expansion ($M$).  Square: $\Delta t = 10^{-2}$; Circle: $\Delta t = 10^{-3}$.  Solid line: $n_s \approx 0.2885M - 3.4684$; dashed line: $n_s \approx 1.5386 M - 3.7472$.   }
\label{figure:accuracy:delta-60}
\end{figure}

\begin{figure}
\centering
\includegraphics[scale=0.3]{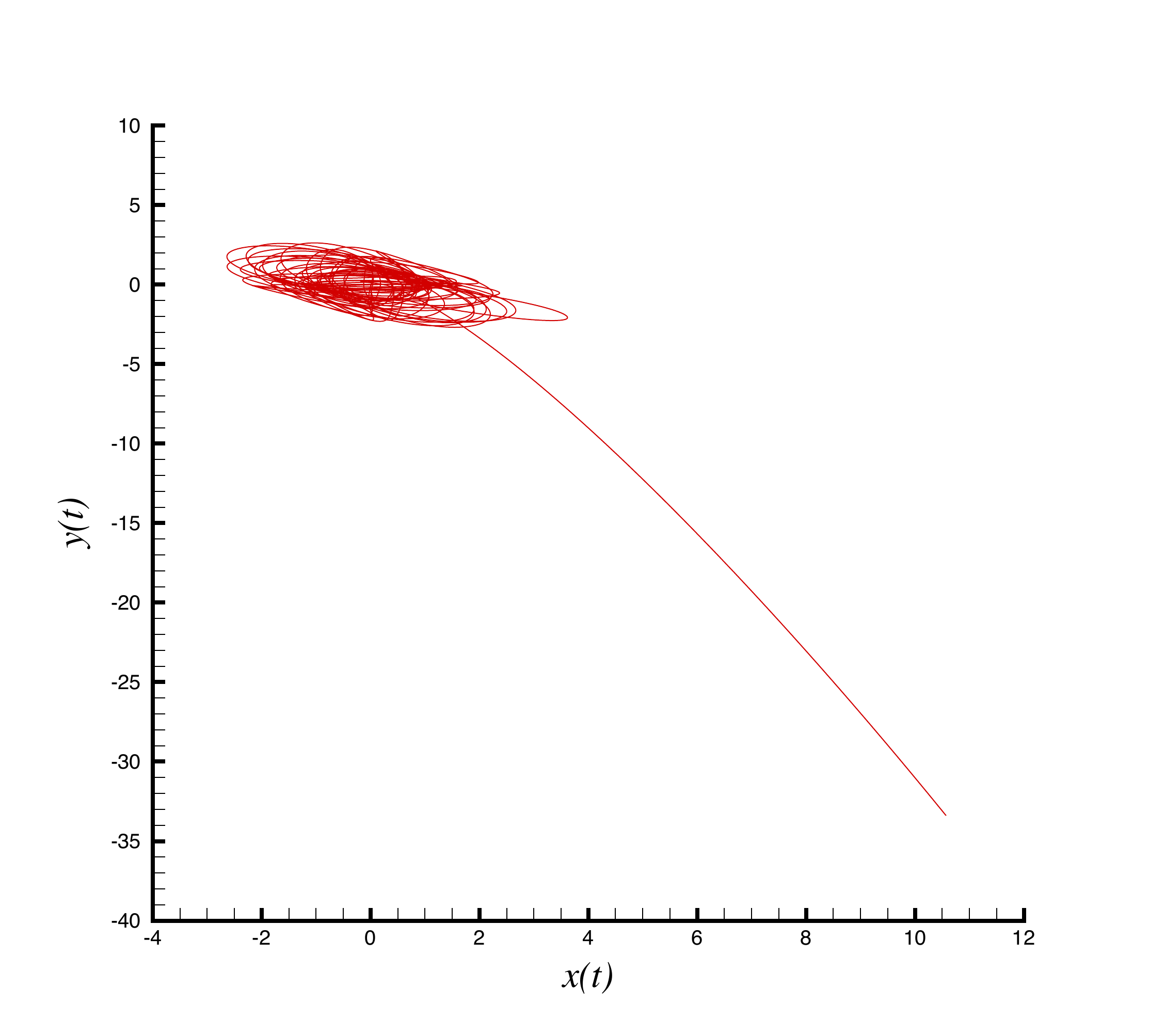}
\includegraphics[scale=0.3]{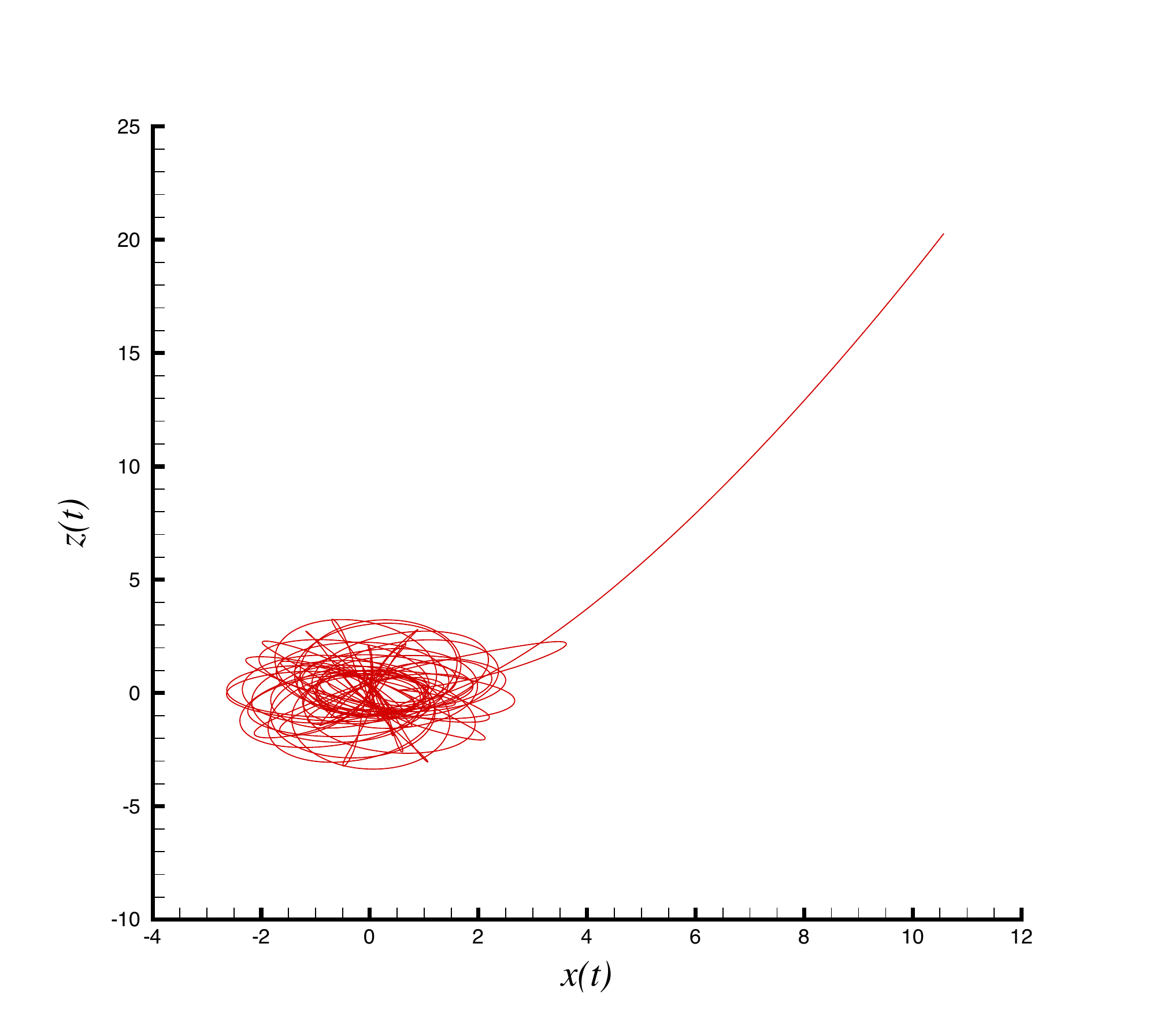}
\caption{ $x-y$ and $x-z$ of Body 1  ($0 \leq t \leq 1000$) in the case of $\delta = 10^{-60}$. }
\label{figure:body1dX-2D}
%\end{figure}

%\begin{figure}
\centering
\includegraphics[scale=0.3]{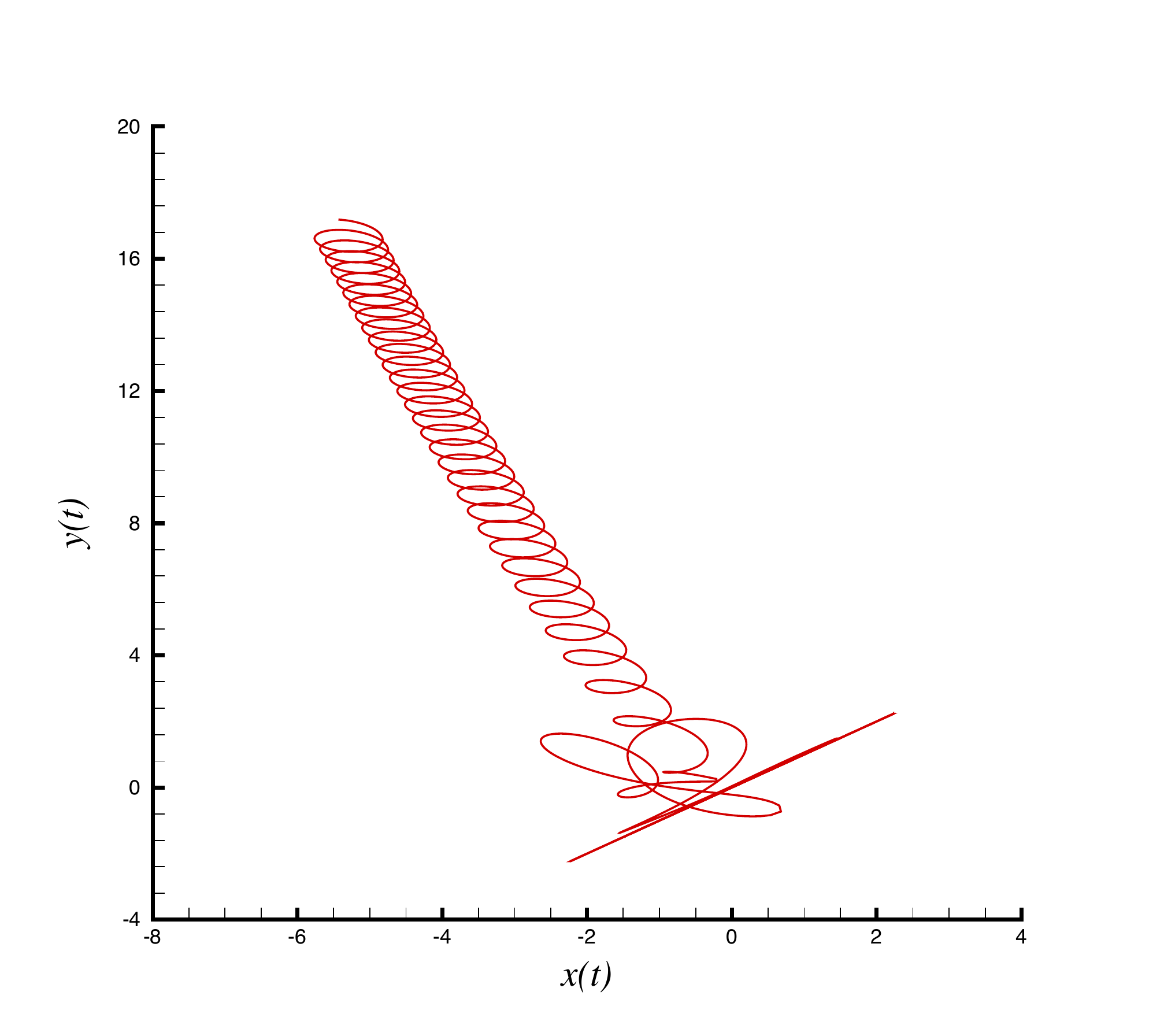}
\includegraphics[scale=0.3]{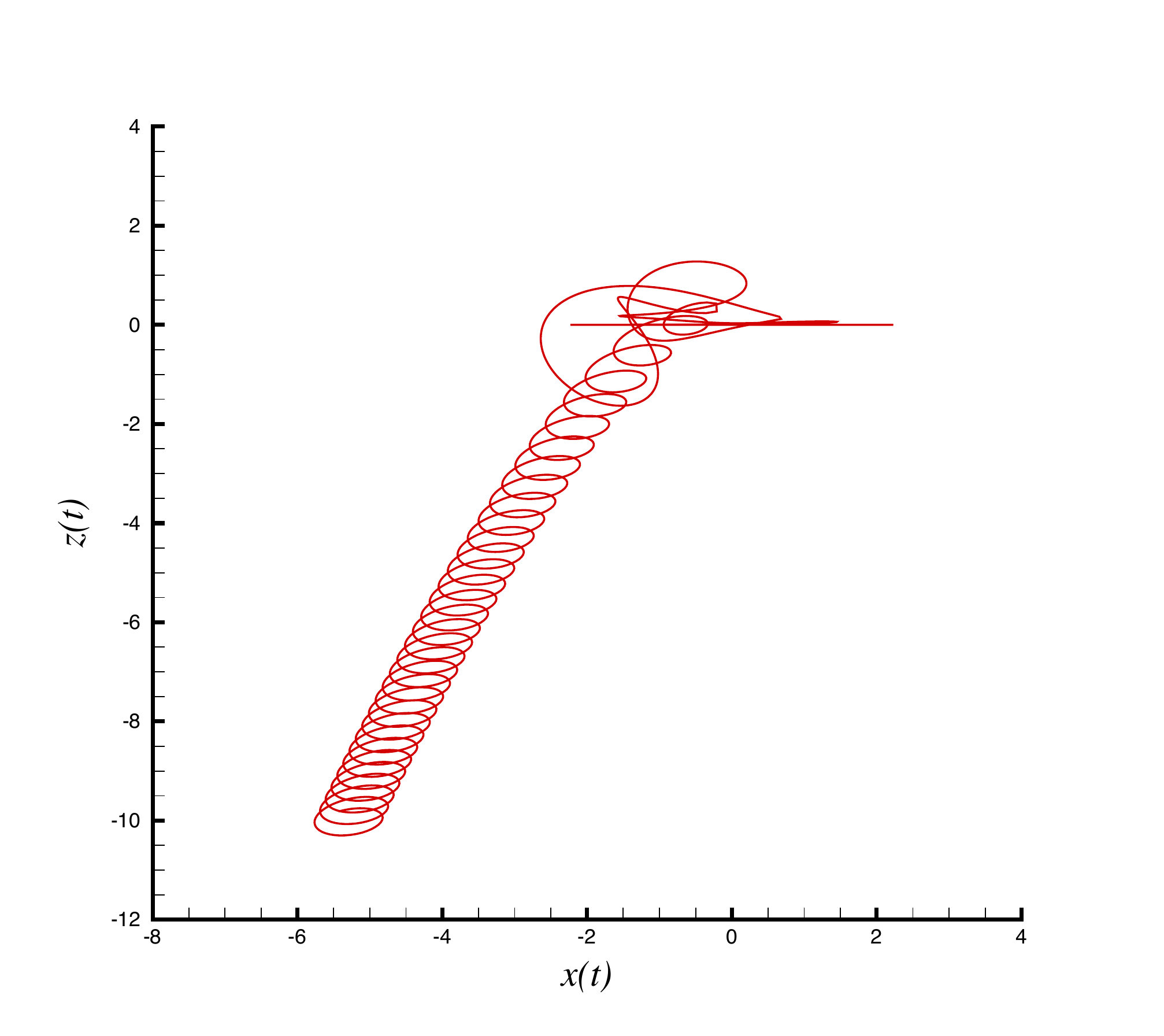}
\caption{ $x-y$ and $x-z$ of Body 2  ($0 \leq t \leq 1000$) in the case of $\delta = 10^{-60}$. }
\label{figure:body2dX-2D}
%\end{figure}

%\begin{figure}
\centering
\includegraphics[scale=0.3]{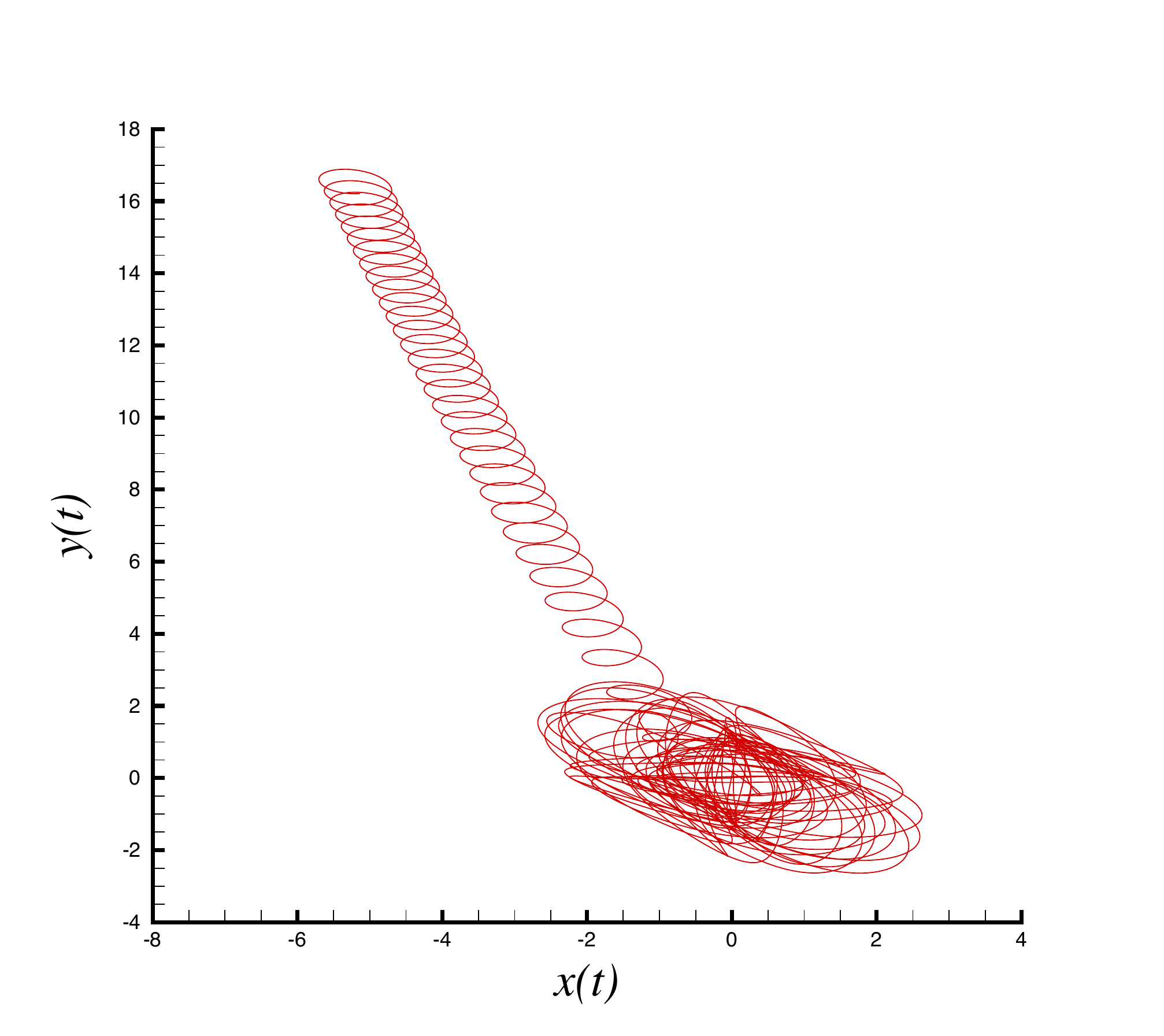}
\includegraphics[scale=0.3]{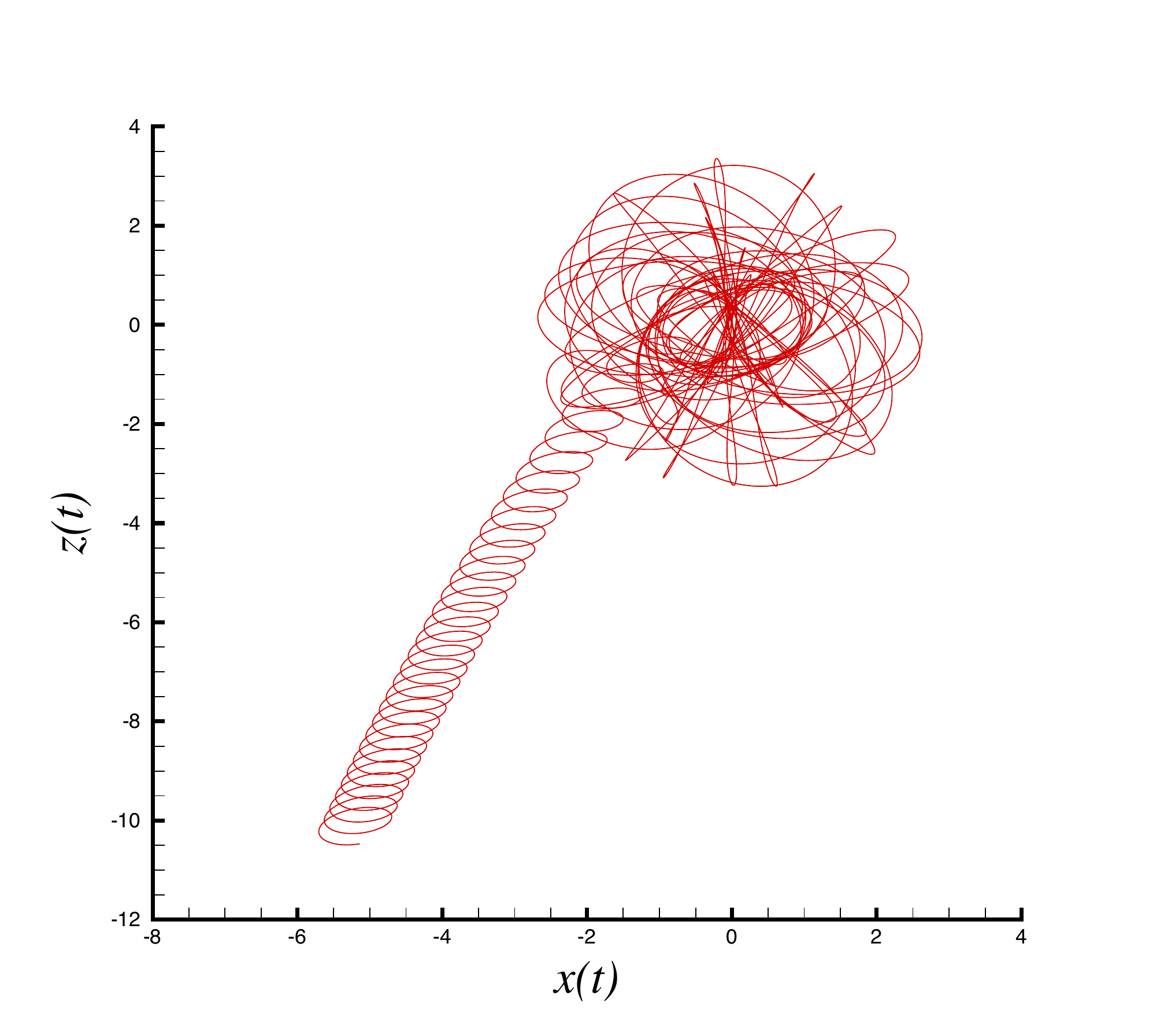}
\caption{ $x-y$ and $x-z$ of Body 3  ($0 \leq t \leq 1000$) in the case of $\delta = 10^{-60}$. }
\label{figure:body3dX-2D}
\end{figure}

\begin{figure}
\centering
\includegraphics[scale=0.3]{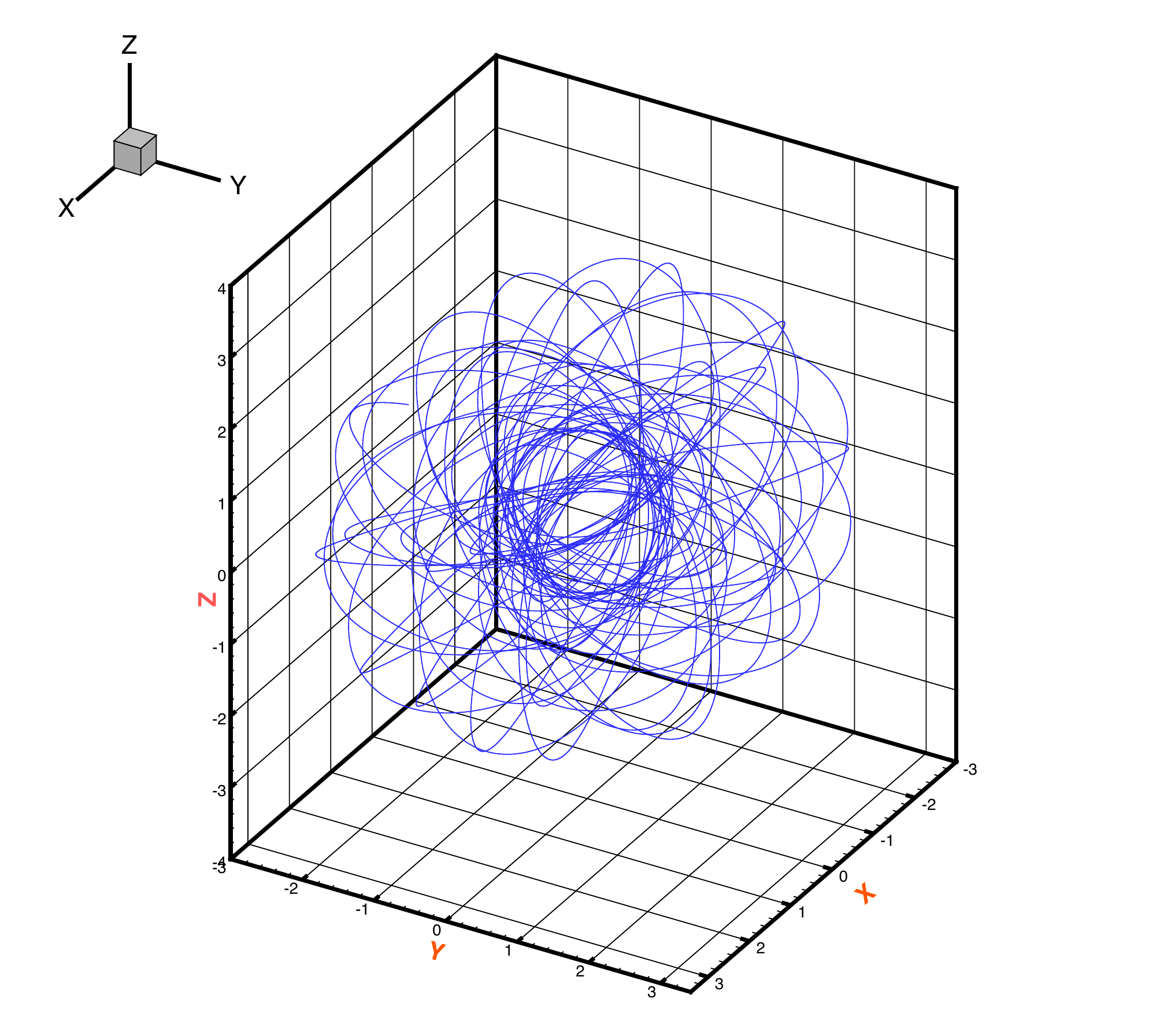}
\includegraphics[scale=0.3]{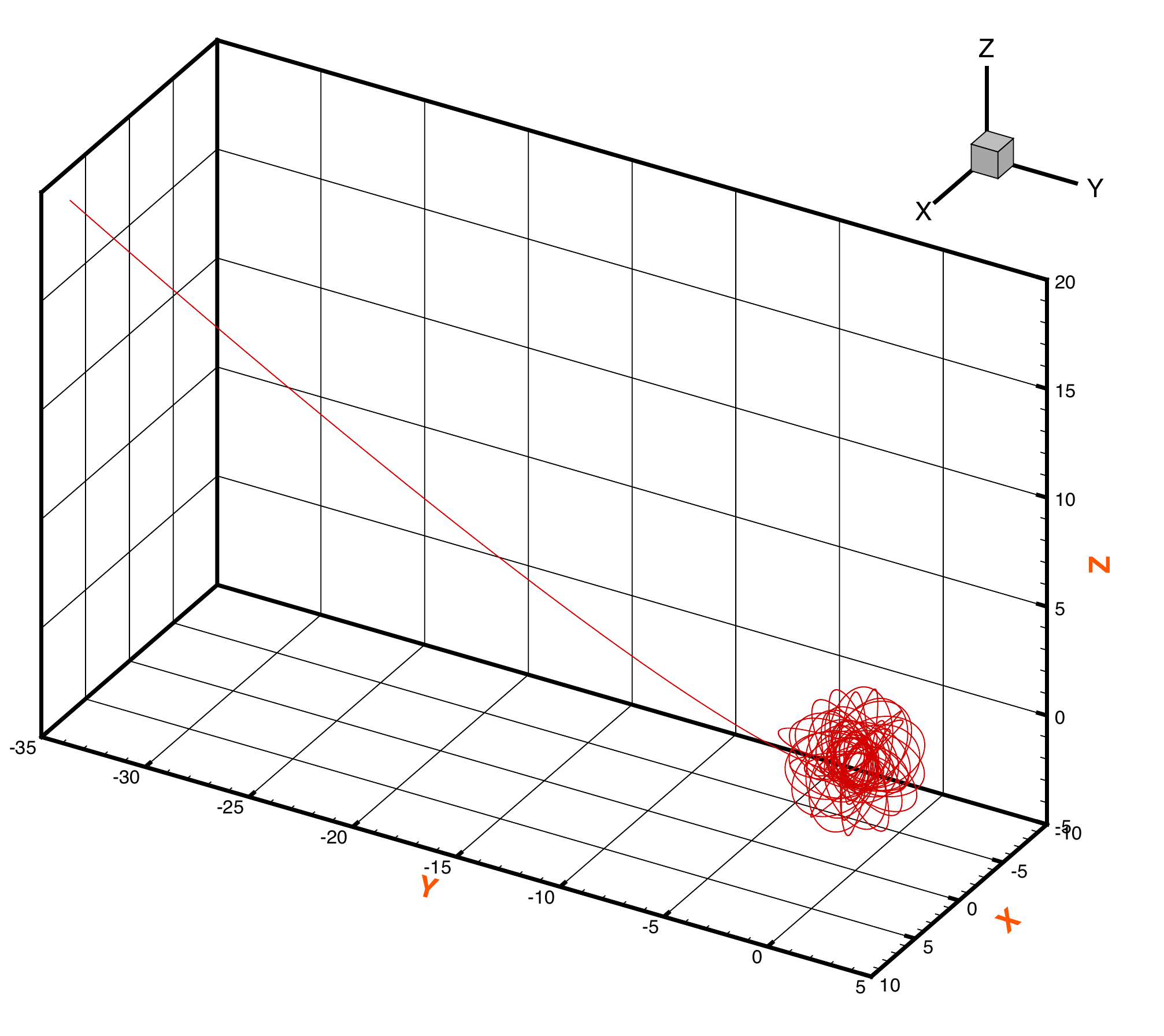}
\caption{  Orbit of Body 1 ($0 \leq t \leq 1000$).  Left: $\delta = 0$;  Right: $\delta = 10^{-60}$. }
\label{figure:body1-3D}
%\end{figure}

%\begin{figure}
\centering
\includegraphics[scale=0.3]{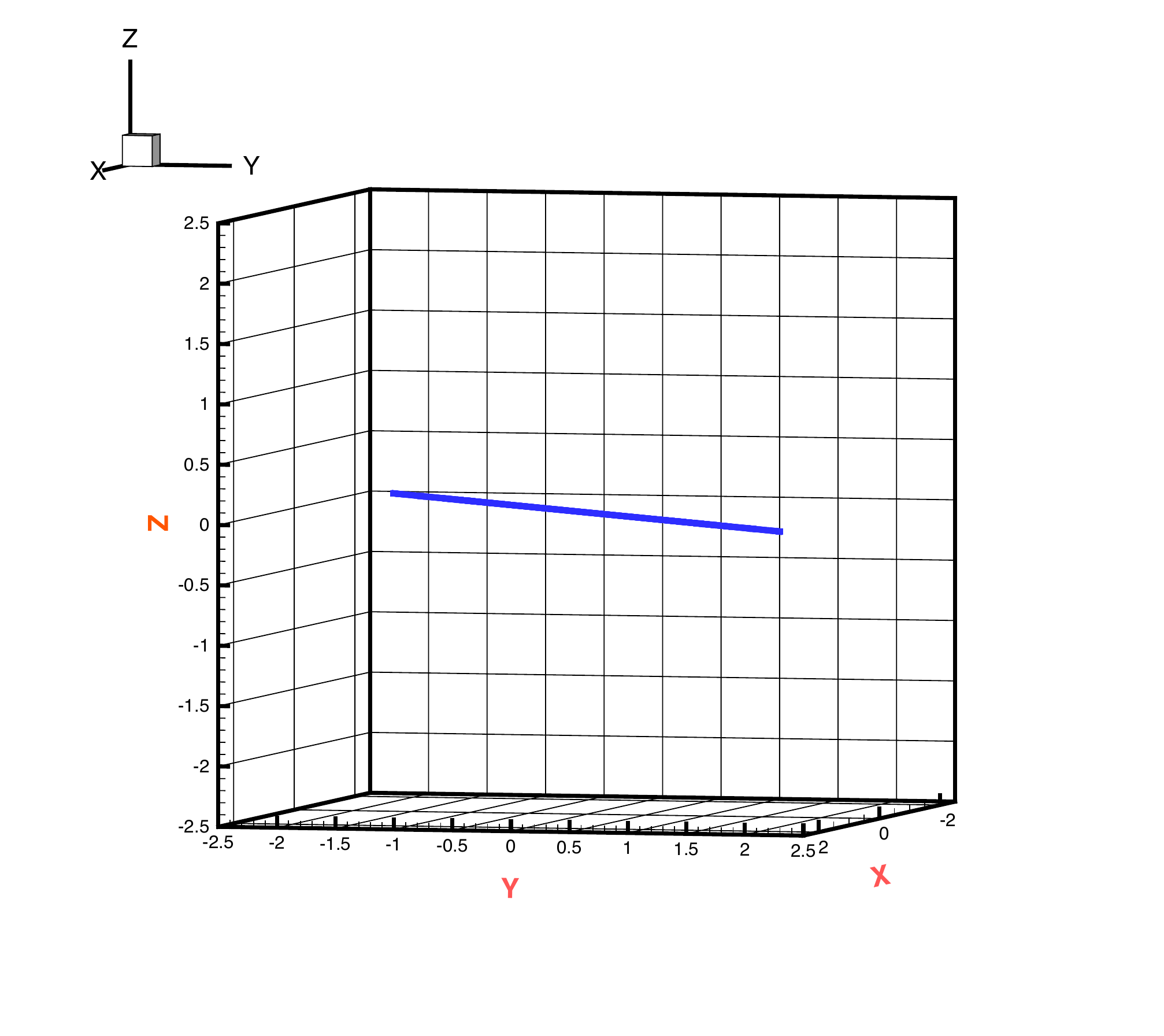}
\includegraphics[scale=0.3]{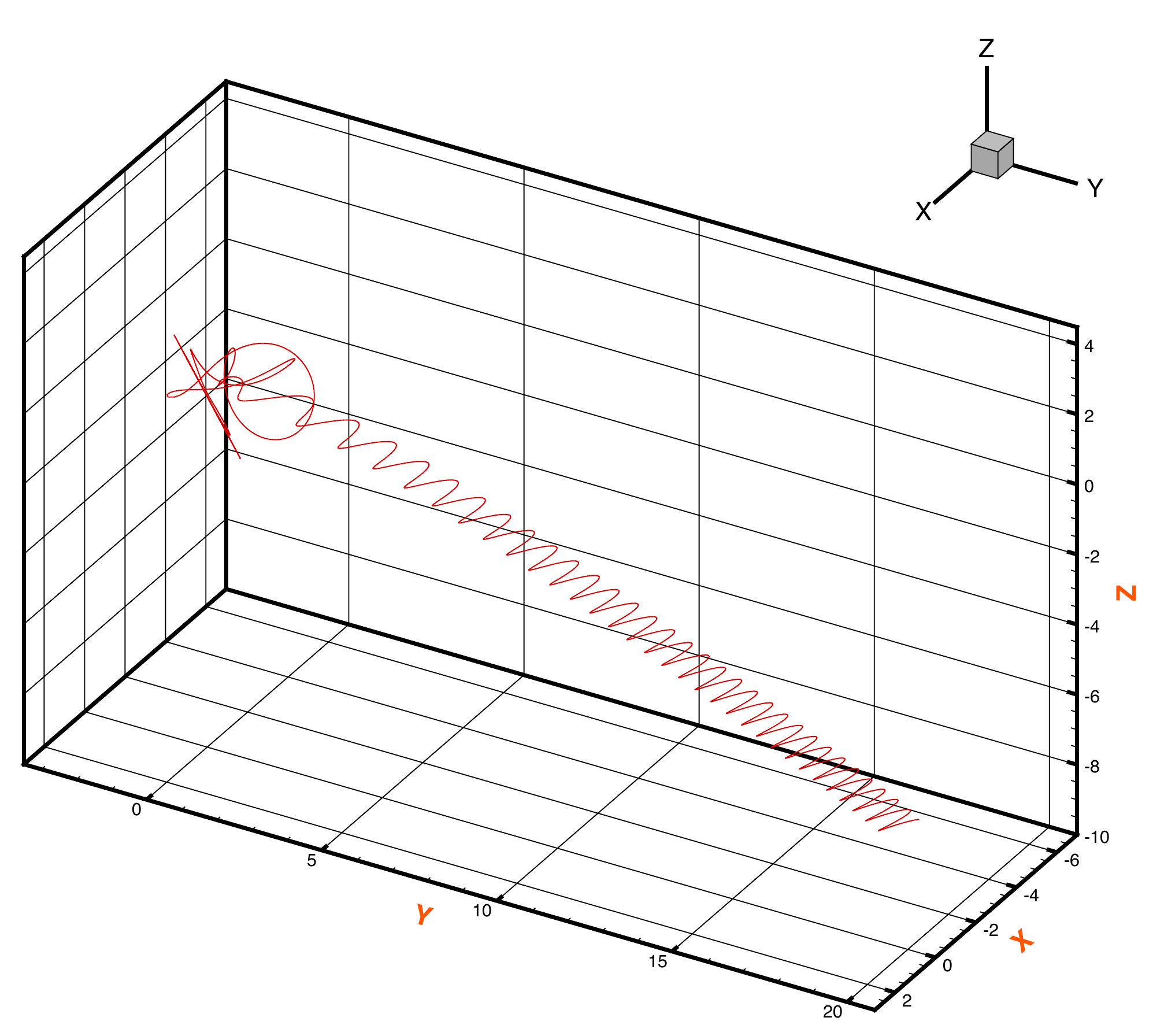}
\caption{  Orbit of Body 2  ($0 \leq t \leq 1000$).   Left: $\delta = 0$;  Right: $\delta = 10^{-60}$.  }
\label{figure:body2-3D}
%\end{figure}

%\begin{figure}
\centering
\includegraphics[scale=0.3]{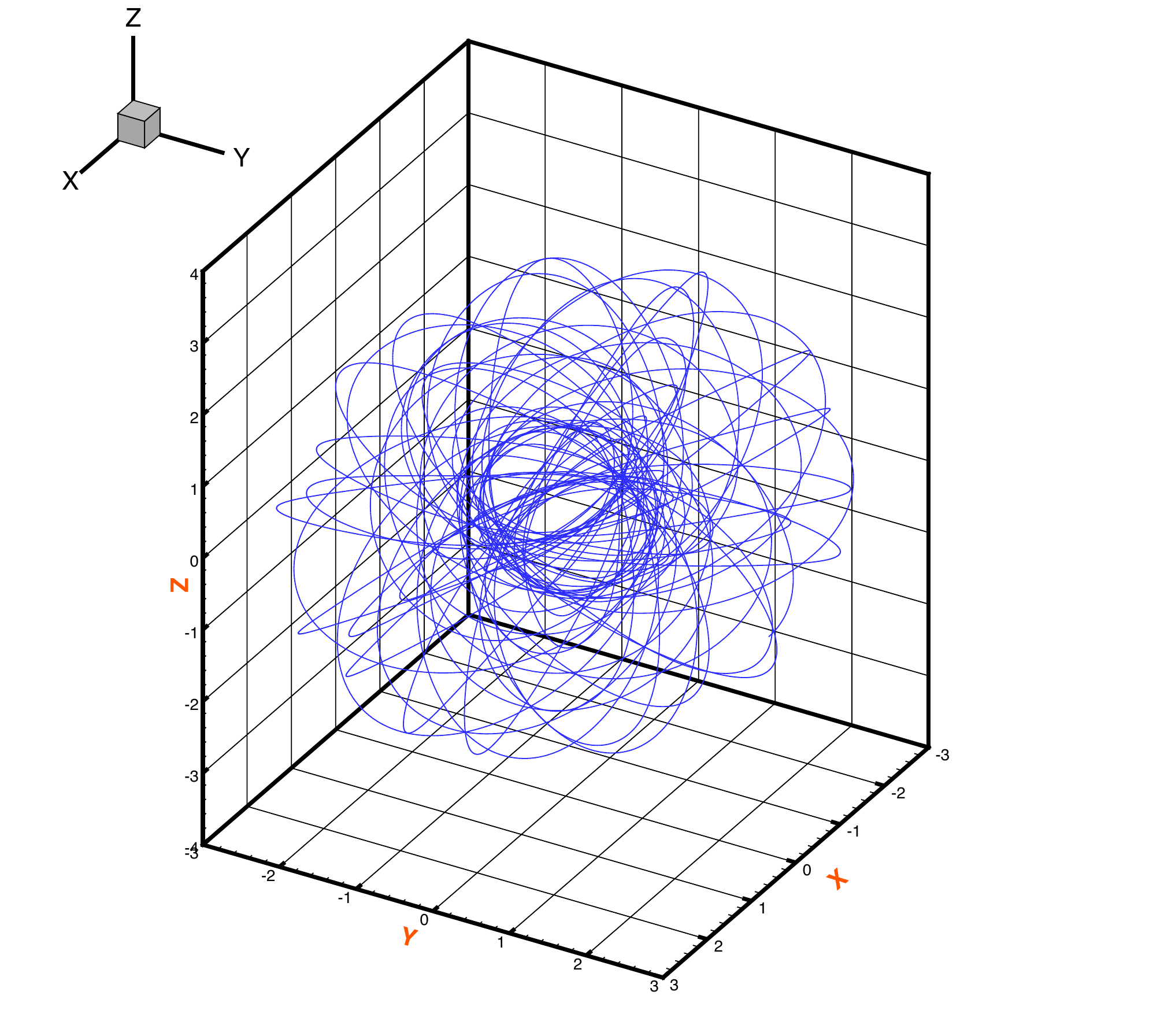}
\includegraphics[scale=0.3]{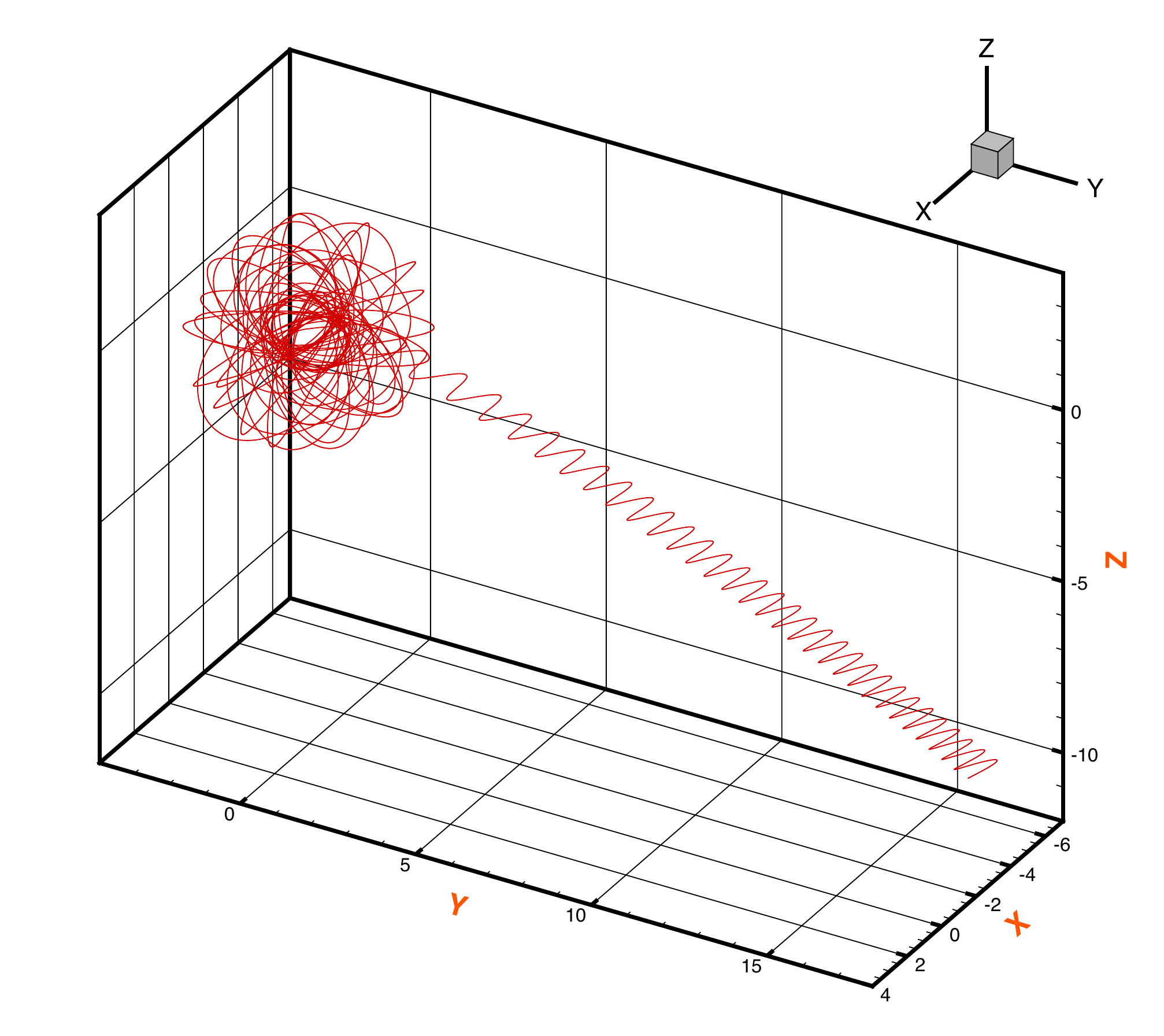}
\caption{  Orbit of Body 3  ($0 \leq t \leq 1000$).   Left: $\delta = 0$;  Right: $\delta = 10^{-60}$.  }
\label{figure:body3-3D}
\end{figure}

The initial  conditions when $\delta=10^{-60}$ have  a tiny physical uncertainty \[  d {\bf r}_1 = 10^{-60} (1,0,0).\]   Since it is very small, it is reasonable to guess that the corresponding dynamic system is chaotic, too.  Similarly, the corresponding  orbits of the three bodies can be  reliably   simulated  by means of the CNS.    It is found that, when $\delta=10^{-60}$,   the CNS simulations by means of $\Delta t = 10^{-2}$, $N=300$  and $M$ = 16, 24, 30, 40, 50, 60 , 70, 80, 100  agree  each other  in the whole interval $[0,1000]$  in the accuracy of  1, 3, 5, 9, 11, 14, 17, 19 and 27  significance digits,  respectively.    Approximately, $n_s$ (the number of significance digits) is linearly  proportional to $M$ (the order of Taylor expansion), say, \[ n_s \approx 0.2885 M - 3.4684,\] as shown in Fig.~\ref{figure:accuracy:delta-60}.   According to this formula, in order to have the CNS simulations (in the interval $[0,1000]$) in the precision of 81 significance digits by means of  $\Delta t=10^{-2}$,  the   300th-order of Taylor expansion, i.e. $M=300$, must be used.    But, this needs much more CPU time.   To  confirm the correction of these CNS simulations,  we further  use  the smaller time step $\Delta t=10^{-3}$.   It is found that,
when $\delta=10^{-60}$ , the CNS simulations  using $\Delta t = 10^{-3}$, $N=300$ and  $M$ = 8, 16, 24, 30,  40, 50 agree well  in the precision of  8,  21, 33, 43, 59, 72 significance digits in the whole interval $[0,1000]$,  respectively.    Approximately, $n_s$, the number of significance digits of the corresponding CNS simulations  in the interval $[0, 1000]$, is linearly  proportional to $M$ (the order of Taylor expansion), say, \[  n_s \approx1.5386 M - 3.7472 \]  as shown in Fig.~\ref{figure:accuracy:delta-60}.
For example,  the  position of Body 1 at $t = 1000 $ given by the 50th-order Taylor expansion and data in 300-digit precision with $\Delta t=10^{-3}$ reads
\begin{eqnarray}
x_{1,1} &=& + 10.57189 91771 62684 86053 99651 18023 33873 55185 816 \nonumber \\
&& 19 03577 63318 20966  52436 81014 64,  \label{x[1,1]-delta-60}\\
x_{2,1} &=& -33.39568 60196 58214 70317 81512 36176 86024 07559 680 \nonumber \\
&& 29 19927 61867 10038 14287 39294 80, \\
x_{3,1} &=& +  20.28455 27396 82192 29521 36869 21793 88441 04404 153 \nonumber \\
&& 94 34152 86710 55848 80509 72622 14,\label{x[3,1]-delta-60}
\end{eqnarray}
which are in the precision of 72 significance digits.    Note  that  the  positions of the  three bodies   at  $t=1000$  given by $\Delta t=10^{-2}$ and $M=100$ agree well (in precision of 27 significance digits) with those by $\Delta t=10^{-3}$ and  $M=50$.    In addition, the momentum conservation (\ref{conversation:r}) is satisfied in the level of $10^{-293}$.    Thus,  our  CNS  simulations  in the case of  $\delta=10^{-60}$  are  {\em mathematically}  reliable  in  the {\em whole}   interval $0\leq t \leq 1000$ as well.

The orbits of the three bodies in the case of $\delta=10^{-60}$ are as shown in Figs.~\ref{figure:body1dX-2D} to \ref{figure:body3dX-2D}.   It is found that, in the time interval $0 \leq t \leq 810$,  the chaotic orbits of the three bodies are not obviously different from those in the case of $\delta = 0$, say, Body~2 oscillates along the same line on $z=0$, Body~1 and Body~3 are chaotic with the same symmetry about the regular orbit of Body~2.   However,  the obvious difference of  orbits  appears when $t\geq 810$: Body 2 departs from the oscillations along the line on $z=0$ and  escapes (together with Body~3) along a complicated three-dimensional  orbit.   Besides, Body~1 and  Body~3  escape  in  the opposite  direction without any symmetry.   As shown in Figs.~\ref{figure:body1-3D} to \ref{figure:body3-3D},  Body~2 and Body~3 go far and far away from Body~1 and thus  become a binary-body system.  It is very interesting that,  when $t > 810$,   the  tiny physical uncertainty $d {\bf r}_1 =10^{-60}(1,0,0)$ of the initial conditions  disrupts  not only the  elegant  symmetry  of  the orbits but also even the three-body system itself!

\begin{figure}
\centering
\includegraphics[scale=0.3]{body1-dX-60.pdf}
\includegraphics[scale=0.3]{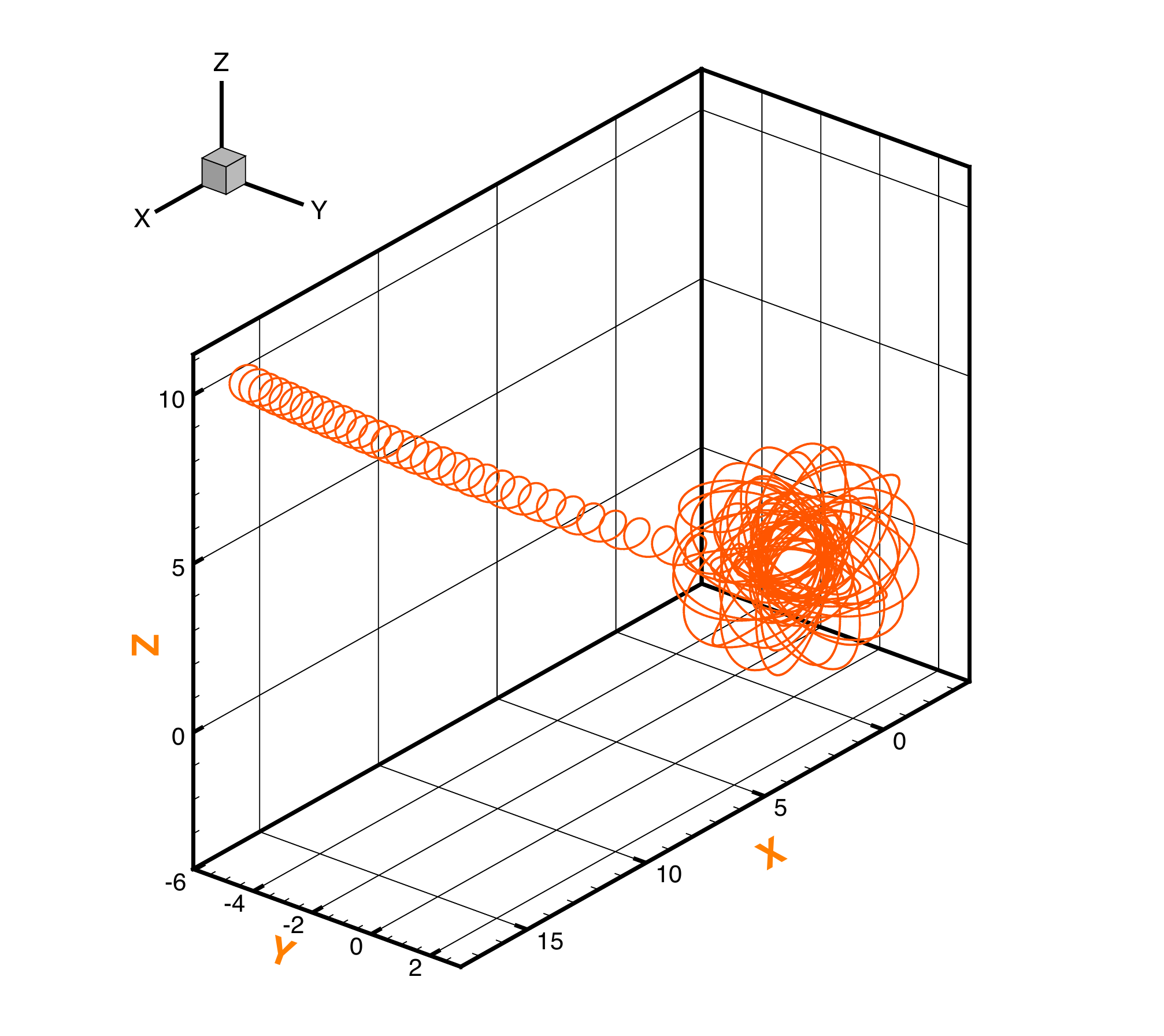}
\caption{  Orbit of Body 1 ($0 \leq t \leq 1000$).  Left: $\delta = +10^{-60}$;  Right: $\delta = -10^{-60}$. }
\label{figure:body1-3D-negative}
%\end{figure}

%\begin{figure}
\centering
\includegraphics[scale=0.3]{body2-dX-60.pdf}
\includegraphics[scale=0.3]{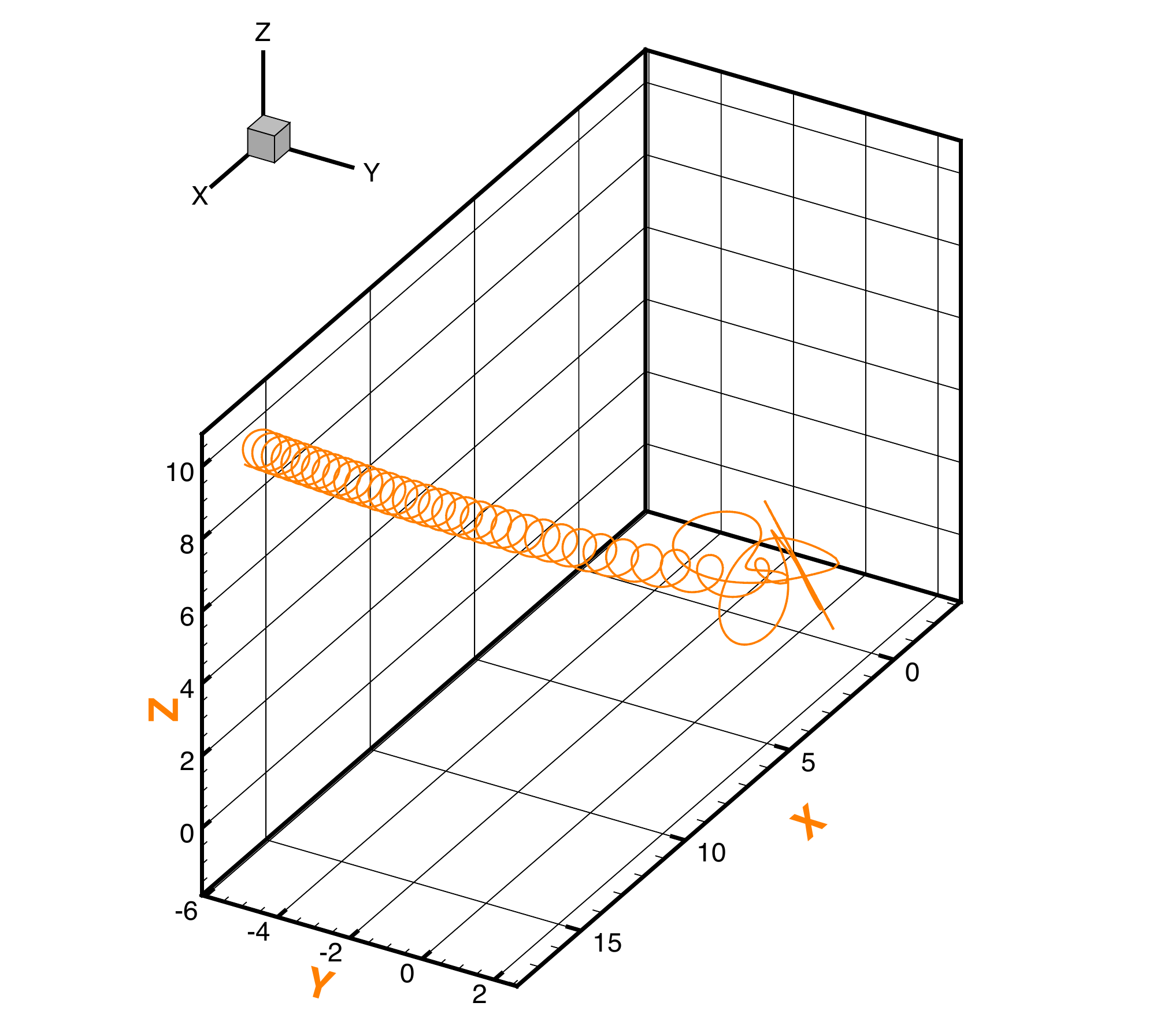}
\caption{  Orbit of Body 2  ($0 \leq t \leq 1000$).   Left: $\delta = +10^{-60}$;  Right: $\delta = -10^{-60}$.  }
\label{figure:body2-3D-negative}
%\end{figure}

%\begin{figure}
\centering
\includegraphics[scale=0.3]{body3-dX-60.pdf}
\includegraphics[scale=0.3]{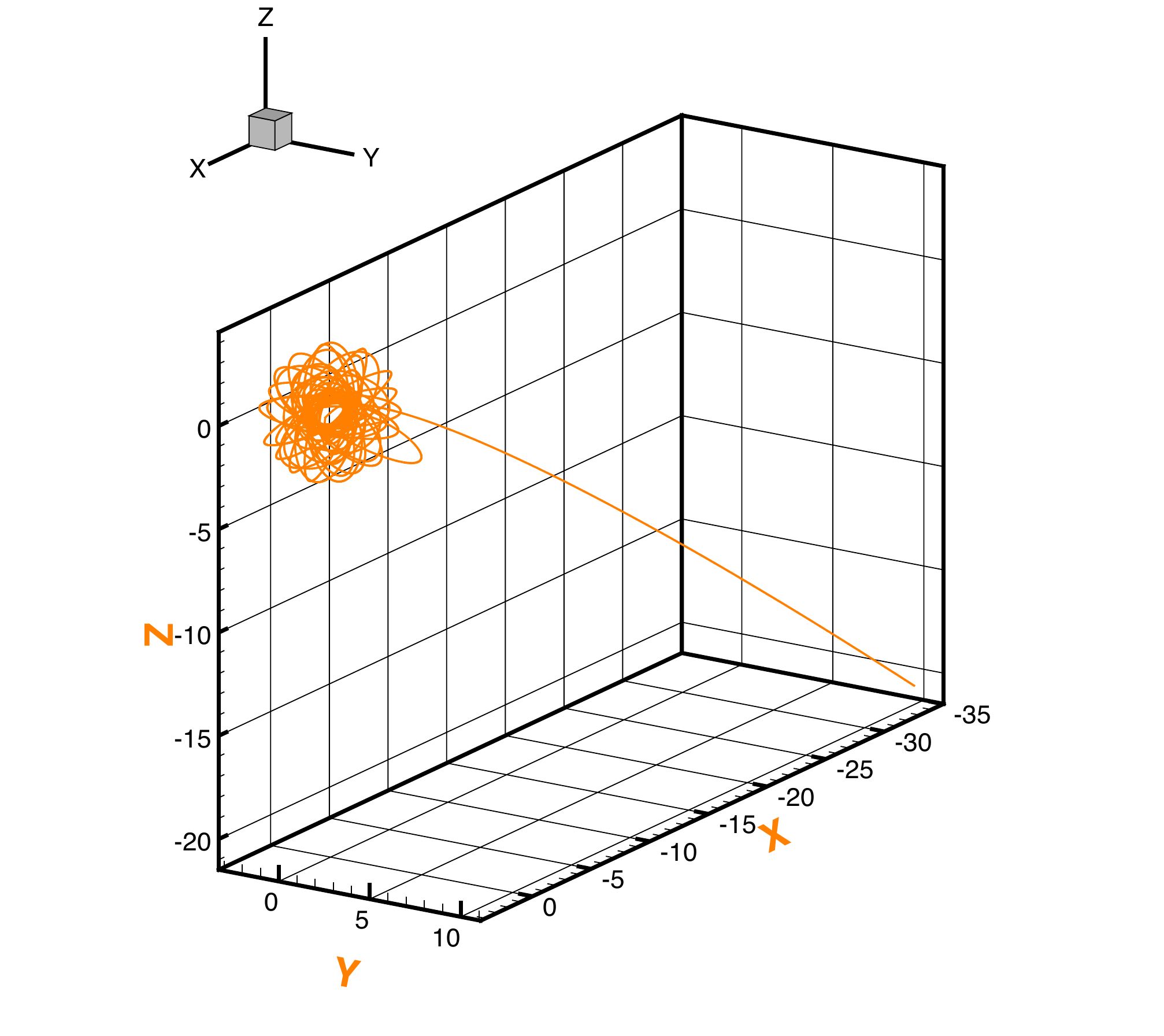}
\caption{  Orbit of Body 3  ($0 \leq t \leq 1000$).   Left: $\delta = +10^{-60}$;  Right: $\delta = -10^{-60}$.  }
\label{figure:body3-3D-negative}
\end{figure}

Similarly,  in the case of $\delta=-10^{-60}$,  we gain the mathematically reliable orbits of the three bodies on the whole interval [0,1000] by means of the CNS with $\Delta t =10^{-3}$, $N=300$ and $M=20$ (i.e. the 20th-order Taylor expansion).   As shown in Figs.~\ref{figure:body1-3D-negative} to \ref{figure:body3-3D-negative},  the tiny physical uncertainty in the initial position disrupts not only the elegant symmetry of the orbits but also the three-body system itself as well:  when $t>810$,  Body~2 departs from its oscillation along the line on $z=0$ and escapes (but together with Body~1) in a complicated three-dimensional orbit, while Body~1 and Body~3  escape  in the opposite  direction   without  any  symmetry.    Note that,  in the case of $\delta=-10^{-60}$,  Body~1 and Body~2   go  together  far and far away  from Body~3  to become a binary system.  However, in the case of $\delta=+10^{-60}$, Body~2 and Body~3 escape together to become a binary system!  This is very interesting.   Thus,  the chaotic orbits of the three-body system subject to the   initial conditions with the micro-level physical uncertainty, corresponding to  $\delta = 0$,  $\delta = +10^{-60}$ and $\delta = -10^{-60}$,  respectively,  are completely {\em different} when $t > 810$.  It should be emphasized here that, these {\em mathematically} different initial conditions are {\em physically} the {\em same}, since any lengths shorter than the Planck length do not make physical senses  \cite{Polchinshi1998}.

Our mathematically reliable CNS simulations  reveal  that there exists such an interval $[0,T_p^{max}]$, where $T_{max}^p \approx 810$ for the special case of the three-body problem, that the chaotic trajectories of the three bodies have no obvious difference\footnote{ In other words, they look like  ``deterministic''} in $[0,T_p^{max}]$, but become  obviously  different beyond it.   It should be emphasized that, by means of the CNS with high enough order of  Taylor series method and data in high enough precision,  the numerical noises are negligible so that all chaotic trajectories are mathematically  reliable in the whole interval $[0,1000]$.   On the other side,  the micro-level physical uncertainty of positions of these initial conditions are shorter even than  the Planck length  so  that  they  are the {\em same} in {\em physics}, according to the string theory \cite{Polchinshi1998}.  In practice,  each of these mathematically different (but physically the same) initial conditions can sample with equal probability,  but we do not know which one will practically occur.  Therefore,  from physical viewpoint,  the chaotic trajectories when $t > T_{max}^p$  are essentially uncertain, i.e. we can not predict the {\em physically} correct trajectories  after  $t > T_{max}^p$, even if they are mathematically reliable.  So, micro-level physical uncertainty of initial condition put forwards a physical limit of prediction time, $T_p^{max}$, which is objective, i.e. independent of numerical methods and  limited precision of measurement  at all.   The key point is that our ability of prediction is greatly restricted by  physical limit of prediction time.   It is a surprise that, for the chaotic motion of the special case of the three body problem considered in this article, the physical limit of prediction time is  a little short, i.e. $T_{max}^p \approx 810$, although it is easy for us to apply the CNS to gain mathematically reliable chaotic trajectories in a much longer interval.     The  the time  of physical limit of prediction  provides  us  a {\em time-scale} for at most how  long  a  deterministic prediction of chaotic dynamic systems is {\em physically} correct.

In addition, the concept of the physical limit of prediction time  suggests that  the  micro-level, objective, physical uncertainty of chaotic systems  might transfer  into  macroscopic uncertainty  for a long time, i.e. after $t > T_{max}^p$.   At least, the considered special case of the three body problem provides us with a good example that the micro-level physical uncertainty might be an origin of some macroscopic uncertainty.  This  suggests  that  macroscopic  uncertainty  might have a close relationship with micro-level uncertainty and thus  is  essentially unavoidable.   This conclusion is supported by the mathematical theory \cite{Wolpert2008}  and  many  physical experiments  (such as those in  \cite{Bai1994, Xia2000}).

\begin{figure}
\centering
\includegraphics[scale=0.3]{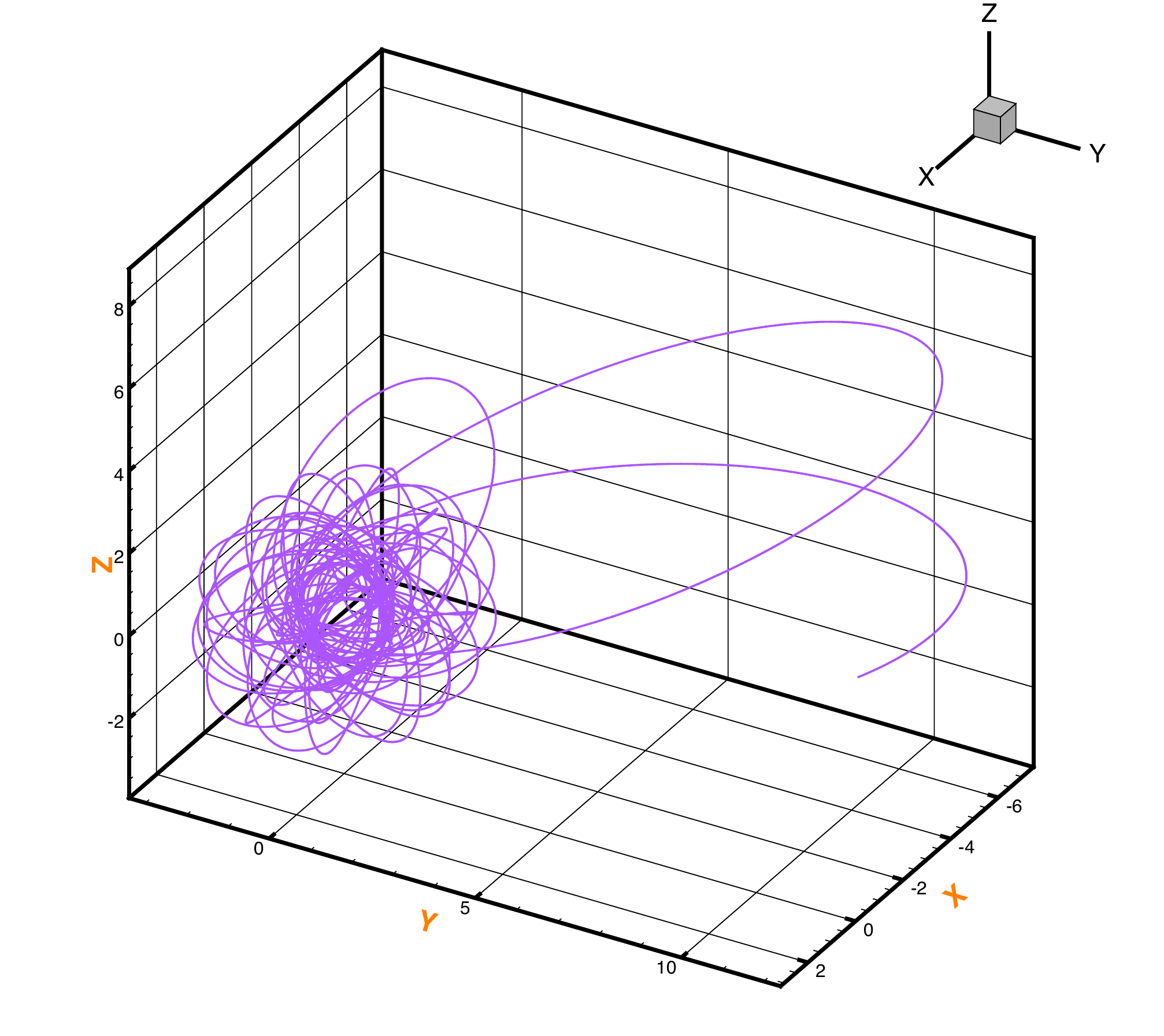}
\caption{  Orbit of Body 1 ($0\leq t\leq 1200$) when $d {\bf r}_1 = 10^{-60}\; \; (1,1,1)$. }
%\caption{  Orbit of Body 1.  Left: $\delta = 0$;  Right: $d {\bf r}_1 = 10^{-60}\; \; (1,1,1)$. }
\label{figure:body1-3D-XYZ}
%\end{figure}

%\begin{figure}
\centering
\includegraphics[scale=0.3]{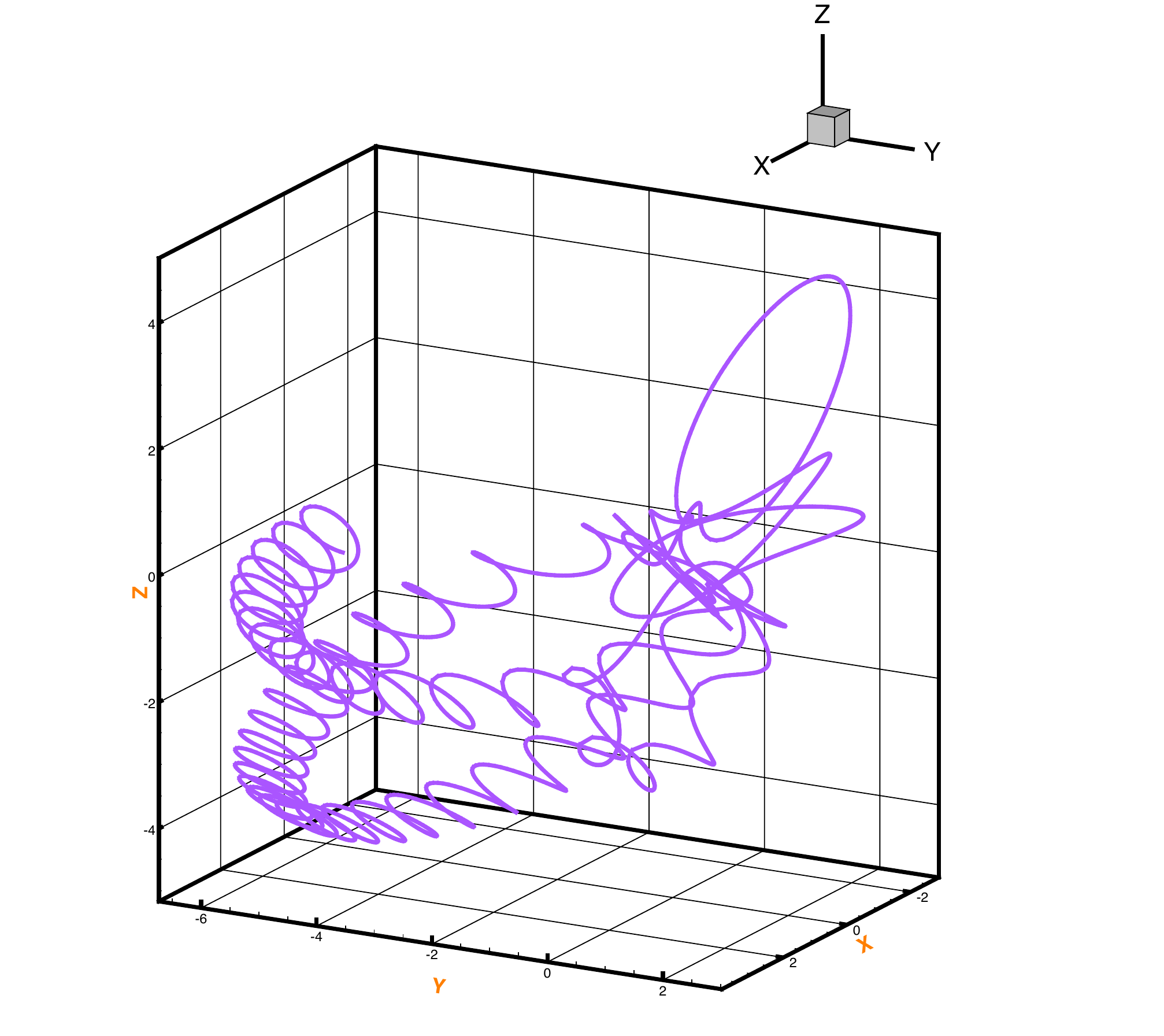}
\caption{  Orbit of Body 2 ($0\leq t\leq 1200$) when $d {\bf r}_1 = 10^{-60}\; \; (1,1,1)$. }
%\caption{  Orbit of Body 2.  Left: $\delta = 0$;  Right: $d {\bf r}_1 = 10^{-60}(1,1,1)$.  }
\label{figure:body2-3D-XYZ}
%\end{figure}

%\begin{figure}
\centering
\includegraphics[scale=0.3]{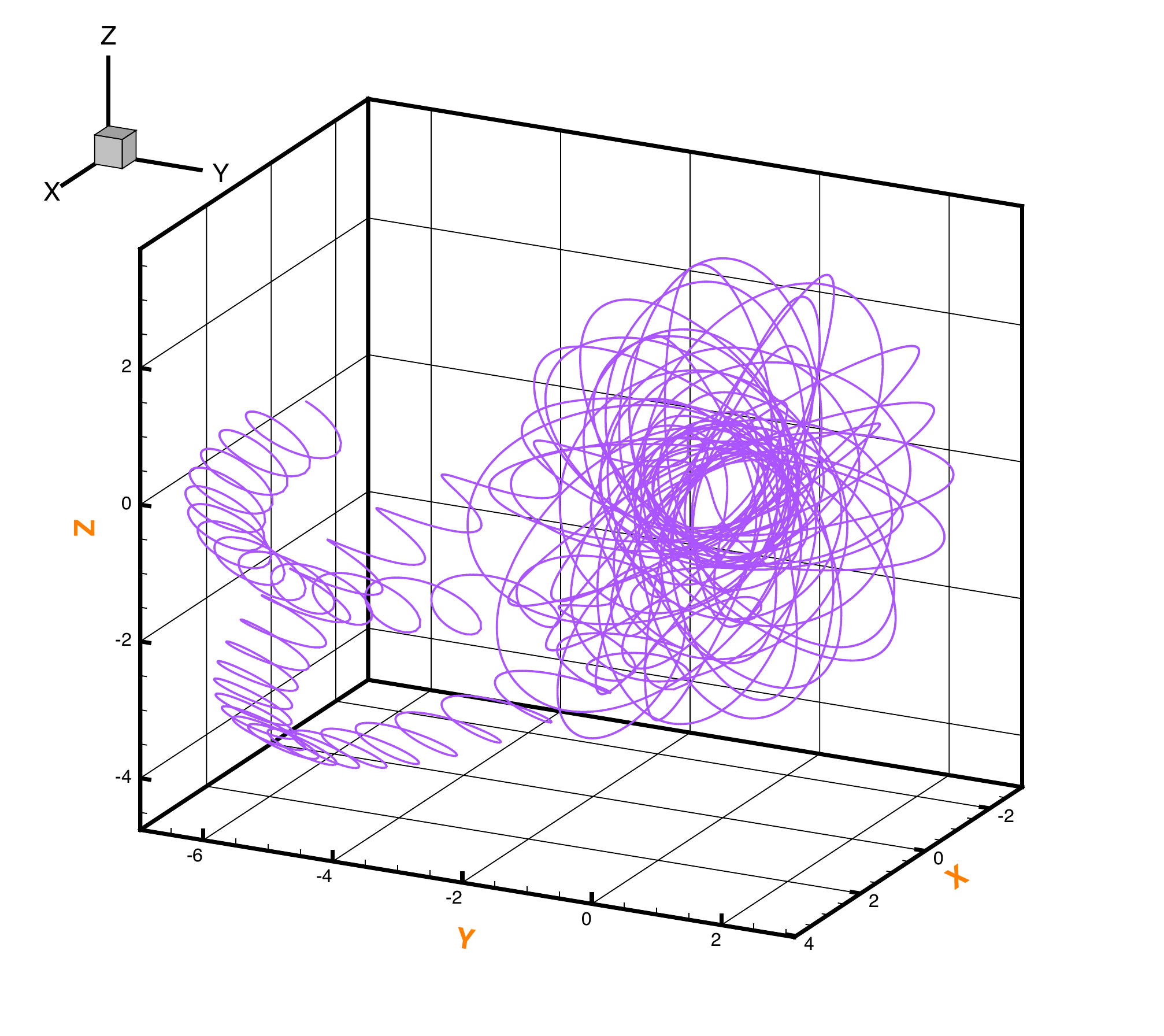}
\caption{  Orbit of Body 3 ($0\leq t\leq 1200$) when $d {\bf r}_1 = 10^{-60}\; \; (1,1,1)$. }
%\caption{  Orbit of Body 3.  Left: $\delta = 0$;  Right: $d {\bf r}_1 = 10^{-60}(1,1,1)$.  }
\label{figure:body3-3D-XYZ}
\end{figure}

Finally, to confirm our above conclusions,  we further consider  such a special case with the micro-level uncertainty of the initial position $d {\bf r}_1 = 10^{-60} (1, 1, 1)$, i.e.
\[   {\bf r}_1 = (0,0,-1) + 10^{-60} (1, 1, 1).  \]
Since the physical uncertainty of position is even shorter than the Planck length,  from the physical viewpoint mentioned above,  the initial positions can be regarded as the {\em same} \cite{Polchinshi1998} as those of the above-mentioned three initial conditions.   However,  the corresponding  mathematically reliable chaotic orbits of the three bodies (in the time interval $0\leq t \leq 1200$) obtained by means of the CNS with $\Delta t =10^{-3}, N= 300$ and the 30th-order Taylor expansion ($M=30$) are almost the {\em same} in the interval  $[0,T_p^{max}]$, where $T_p^{max}\approx 810$,   but beyond it  they become {\em quite} different from those of the above-mentioned three cases:  Body~2 first oscillates along a line on $z=0$  but  departs  from the regular orbit   when $t > 810$  to move along a complicated three-dimensional orbits, while Body~1 and Body~3 first move with the symmetry but lose it when $t>810$, as shown in Figs.~\ref{figure:body1-3D-XYZ} to \ref{figure:body3-3D-XYZ}.   However, it is not clear whether any one of  them might escape or not,  i.e.  the fate of the three-body system is uncertain.    Since  such  kind  of  micro-level uncertainty of initial position is inherent and objective,  the orbits of the three-body system beyond the time of the physical limit of prediction is essentially uncertain in physics.   It should be emphasized that the uncertainty of chaotic trajectories beyond the  time $T_p^{max}$ of physical limit of prediction is {\em objective}, i.e. it has nothing to do  with  limited precision of measurement,   numerical noises and Heisenberg uncertainty principle \cite{Heisenberg1927}.    All of these confirm once again our conclusions mentioned above.

 \section{Concluding remarks and discussions}

A half century ago, Lorenz \cite{Lorenz1963} made a famous conclusion that ``long-term prediction of chaos is impossible''.    However,  what is the exact meaning of the ``long-term'' in the above famous claim? Is  one day  long enough?    Or one millions of years?

In this article,  using chaotic motion of the famous three body problem as an example,  a new concept, namely the physical limit of prediction time, is put forwarded to provide us a {\em time-scale} for at most how long a {\em mathematically} reliable simulation of chaotic dynamic system is deterministic and {\em physically} correct.   First of all, from mathematical viewpoint, given an exact initial condition, we can always gain reliable simulations of chaotic trajectories of the considered three body problem in a finite but long enough interval (such as $[0,1000]$) by means of  the so-called ``Clean Numerical Simulation'' (CNS) \cite{Liao2009, Liao2013}.   This is mainly because, based on the high enough order of Taylor series method and the data in high enough multiple precision, both of truncation and round-off errors of the CNS simulations are negligible in the finite but long enough interval.   However, from physical viewpoint, the initial position of any a star/planet contains micro-level physical uncertainty, that is objective and unavoidable.  Our mathematically reliable CNS simulations of chaotic trajectories of the considered three-body problem with the physical uncertainty  at the micro-level $10^{-60}$ indicate that there exists such a time of physical limit of prediction, i.e. $T_p^{max}$,  that  the chaotic trajectories of the three body problem  are the same  in the interval $[0,T_p^{max}]$,  but  beyond  it, they become  essentially  uncertain  in physics.   Thus,  the time $T_p^{max}$ of the physical limit of prediction provides us a time-scale of {\em physically} correct prediction for chaos:  it has no physical sense to talk about the ``prediction'' of chaotic trajectories beyond $T_p^{max}$,  since they are essentially uncertain in physics.  The concept of ``the physical limit of prediction'' also implies that the macroscopic model of the three body problem based on Newtonian gravitational law is essentially {\em not} deterministic, because the initial condition contains the micro-level physical uncertainty that is unavoidable and might   become very important for the time beyond the physical limit of prediction, i.e. $T_p^{max}$.

On the other side, the concept of  ``physical limit of prediction''  also  suggests  that  the micro-level physical uncertainty of initial conditions might transfer into macroscopic uncertainty  when $t  > T_p^{max}$.   In other word,  our mathematically reliable CNS simulations of the chaotic motion of the  three body problem under consideration
suggest  that  the seemingly random distribution of stars/planets in the sky contains  uncertainty,  and that the micro-level physical uncertainty of position of a star/planet might be an origin of this kind of macroscopic uncertainty.   So, it might provide us a theoretical explanation to many questions, such as  why stars in the sky look random, why velocities of turbulent flows are so uncertain, and so on.   All  of  these  might enrich our knowledge and deepen our understandings about not only the three-body problem but also chaos.  Besides, it also illustrates that the CNS  might provide us a safe numerical method to understand the physical world better.

A special  three body problem, subject to the four mathematically different initial conditions with micro-level physical uncertainty,  is considered in this paper.  The four initial conditions contains a tiny difference due to the physical uncertainty of position in the micro-level of $10^{-60}$: they are shorter even than  the Planck length  so  that  they  are the {\em same} in {\em physics}, according to the string theory \cite{Polchinshi1998}.   In practice,  each of these {\em mathematically} different (but {\em physically} the same) initial conditions can sample with equal probability,  but we do not know which one practically occurs.  Using the CNS, we gain mathematically reliable chaotic trajectories in the interval [0,1000] for each of these four (mathematically different but physically same) initial conditions, and found that the corresponding  chaotic trajectories agree well in the interval $[0, 810]$,  but  become  quite  different when $t  > 810$.  So, we have the time $T_p^{max} \approx  810$ of the physical limit of prediction, which provides us a time-scale for at most how long we can gain a {\em physically} correct, deterministic prediction of chaotic motion for the special three-body problem.    Besides,  the  time  $T_p^{max}$ of the physical limit of prediction also provides us  a  better understanding about the ``long-term'' prediction in Lorenz's famous claim \cite{Lorenz1963}.   Finally,  it should be emphasized that,    given an {\em exact} initial condition,  we can gain {\em mathematically} reliable simulations of chaotic dynamic system under consideration in a {\em larger} interval $[0,1000]$ by means of the CNS.  However, such kind of {\em exact} initial condition does {\em not} exist from the physical viewpoint.  It is the interaction of this kind of uncertainty of the initial condition and the so-called butterfly effect of chaotic dynamic system that leads to the physical limit of prediction of chaos.

Note that the computation ability of  human  being  plays an important role in the development of nonlinear  dynamic systems.  Without digit computer, it  was  impossible for Lorenz \cite{Lorenz1963}  to find out the famous  butter-fly effect of chaos, although only 16-digits precision  was  able  to   use  in  his  pioneering work \cite{Lorenz1963}.   Thanks to the great development of computer technology in the past half century,   the CNS  based on arbitrary order of Taylor series method and arbitrary precision of data  provides us a {\em time-scale}, i.e. the time $T_p^{max}$ of the physical limit of prediction, about ``how long''  we can give a {\em physically}  correct, accurate, deterministic prediction of chaotic dynamic systems.

As reported by Sussman and Wisdom \cite{Sussman1988, Sussman1992},  the motion of Pluto and even the solar system is chaotic  with a time scale in the range  of  3  to 30 million years.   Thus,   our illustrative CNS  simulations reported in this paper suggest  that  the solar system might be essentially uncertain beyond a time of physical limit of prediction which might be millions of years.   Note that such kind of macroscopic uncertainty is closely related to the  inherent  physical micro-level uncertainty of position and thus is objective.  In other words, it has nothing to do with limited precision of  measurement:  the history of human being  is  indeed  too short,  compared to the time scale  of such kind of macroscopic uncertainty.

The proposed reliable numerical approach can be used to investigate the stability of periodic solutions of nonlinear dynamic systems.  For example, it can be used to study the stability of the newly found three classes of Newtonian three-body planar periodic orbits \cite{3bodyPRL-2013}.

Note that the time of physical limit of perdition might be quite different for different chaotic systems, different initial conditions,  different physical parameters and so on.  It would be  interesting to investigate the relationship between the time of physical limit of prediction and some traditional characteristics of chaos, such as Lyapunov exponent, and so on.

Finally, it should be emphasized that the microscopic and macroscopic phenomena are described by two completely different systems of physical laws, i.e. the quantum mechanics \cite{quantum} and classical (Newtonian) mechanics, respectively.   It is widely believed that the microscopic phenomena are uncertain and described by the quantum mechanics, but most macroscopic phenomena are governed by deterministic physical laws of classical mechanics (such as the Newtonian second law and the Newtonian gravitational law for three body problem).   Many scientists  attempted to develop a unified theory valid for both of the microscopic and macroscopic phenomena.  Unfortunately, such a unified theory dos not exist up to now.  It is not very clear whether or not and how the micro-level uncertainty has relationships with the macroscopic uncertainty.  However, it  is  a  fact  that  macroscopic  phenomena  with  uncertainty exist widely, such as the distribution of stars in the universe, turbulent flows and so on.  The origin of macroscopic uncertainty of these phenomena and especially their relationships to the micro-level uncertainty are not very clear.  Our current work reported in this article  strongly suggests that  the micro-level physical uncertainty might be an origin of some macroscopic phenomena with uncertainty, although it might be not the only one.  In other words,  our work suggests that the macroscopic uncertainty might have a close relationship with the micro-level uncertainty,  although we have not a unified theory now.  Note that the nonlinearity is an inherent property of the nature and the uncertainty is a fundamental property of microscopic phenomena.  Thus,  our work suggests that the macroscopic uncertainty seems to be inescapable: it  should be  an inherent property of the nature, too.  If this is indeed true, it could well explain the origin of uncertainty of many macroscopic phenomena, such as turbulent flows, the random distribution of stars in the universe and so on.

 \section*{Acknowledgement} This work is partly supported by the State Key Lab of Ocean Engineering (Approval No. GKZD010056-6) and  the National Natural Science Foundation of China under Grant No. 11272209.

 %\newpage


\begin{thebibliography}{99.}

\bibitem{Wolpert2008}
Wolpert, D.H.: ``Physical limits of inference''.  {\em Physica D}, \textbf{237}: 1257-1281 (2008).

\bibitem{Poincare1890}
Poincar\'{e}, J.H.:  ``Sur le probl\`{e}me des trois corps et les \'{e}quations de la dynamique.  Divergence des s\'{e}ries de M. Lindstedt''.  {\em Acta Mathematica}, \textbf{13}:1 -- 270 (1890).


\bibitem{Lorenz1963}
Lorenz, E.N.:  ``Deterministic non-periodic flow''.  {\em Journal of the Atmospheric Sciences}, \textbf{20}: 130 -- 141 (1963).

\bibitem{Lorenz1989}
E.N. Lorenz, ``Computational chaos - a prelude to computational instability''. {\em Physica D}, {\bf 15}: 299 -- 317 (1989).

\bibitem{Lorenz2006}
Lorenz,  E.N.: ``Computational periodicity as observed in a simple system''. {\em Tellus-A},  \textbf{58}: 549 -- 59 (2006).


\bibitem{Li2000}
J.P. Li, Q.G. Zeng and J.F. Chou, ``Computational uncertainty principle in nonlinear ordinary differential equations (I): numerical results'', {\em Science in China (Series E)}, {\bf 43}: 449-460 (2000).

\bibitem{Li2001}
J.P. Li, Q.G. Zeng and J.F. Chou, ``Computational uncertainty principle in nonlinear ordinary differential equations (II): theretical analysis'', {\em Science in China (Series E)}, {\bf 44}: 55-74 (2001).


\bibitem{Teixeira2007}
 J. Teixeira, C.A. Reynolds, and K. Judd,  ``Time Step Sensitivity of Nonlinear Atmospheric Models: Numerical Convergence,
Truncation Error Growth, and Ensemble Design'', {\em J. Atmos. Sci.},  {\bf 64}: 175 -- 188 (2007).

\bibitem{Liao2009}
  Liao, S.J.:   ``On the reliability of computed chaotic solutions of non-linear differential equations''.   {\em Tellus-A}, \textbf{61}: 550 -- 564 (2009).

\bibitem{Liao2012}
Liao, S.J.: ``Chaos: a bridge from microscopic uncertainty to macroscopic randomness''.  {\em Commun. Nonlinear Sci. Numer. Simulat.}, {\bf 17}:   2564 -- 2569  (2012).

\bibitem{Anosov1967}{Anosov1967}
D.V. Anosov, ``Geodesic flows on closed Riemannian manifolds with negative curvature'', {\em Proc. Steklov. Inst. Math.}, {\bf 90}:1 (1967)

 \bibitem{Yorke1994}
S. Dawson, C. Grebogi, T. Sauer, and J.A. Yorke, ``Obstructions to shadowing when a Lyapunov exponent fluctuates about zero'', {\em Phys. Rev. Lett.},  {\bf 73}: 1927-- 1930 (1994).

\bibitem{Yorke1997}
T. Sauer, C. Grebogi, and J. A.  Yorke,  ``How long do numerical chaotic solutions remain valid ?'',  {\em Phys. Rev. Lett.},  {\bf 79}, 59 -- 62  (1997).

  \bibitem{Sauer2002}
T. Sauer, ``Shadowing breakdown and large errors in dynamical simulations of physical systems'', {\em Phys. Rev. E},  {\bf 65}: 036220 (2002)


\bibitem{Yorke2000}
G. Yuan and J.A. Yorke, ``Collapsing of chaos in one dimensional maps'', {\em Physica D}, {\bf 136}: 18 (2000)

\bibitem{Shi2008}
P.L. Shi, ``A relation on round-off error, attractor size and its dynamics in driven or
coupled logistic map system'',  {\em Chaos}, {\bf 18}: 013122 - 8  (2008).

\bibitem{Liao2013}
Liao, S.J.: ``On the  numerical simulation of propagation of micro-level uncertainty for chaotic dynamic systems''. {\em Chaos, Solitons and Fractals}, {\bf 47}: 1 -- 12 (2013).

\bibitem{Corliss1982}
G.F. Corliss and Y.F. Chang, ``Solving ordinary differential equations using Taylor series'',  {\em ACM Trans. Math. Software}, {\bf 8}, 114 --144 (1982).

\bibitem{Barrio2005}
R. Barrio, F. Blesa, and M. Lara, ``VSVO formulation of the Taylor method
for the numerical solution of ODEs'',   {\em Computers and Math. with Applications}, {\bf 50}, 93  --111 (2005).

 \bibitem{MP}
Oyanarte,~P.:  ``MP -- a multiple precision package''. {\em Comput. Phys. Commun},  {\bf 59}: 345 -- 358 (1990).

\bibitem{Wang2011}
  Wang,~P.F.,  Li,~J.P. and Li,~Q.:  ``Computational uncertainty and the application of a high-performance multiple
precision scheme to obtaining the correct reference solution of Lorenz equations'',  {\em Numerical Algorithms},  {\bf 59}: 147 -- 159 (2012).

\bibitem{Liao-Wang}
Liao, S.J. and Wang, P.F.: ``On the reliable long-term simulation of chaos of Lorenz equation in the interval [0,10000]''.  ArXiv:1305.4222 ({http://arxiv.org/abs/1305.4222}).

\bibitem{Diacu1996}
 Diacu, F. and Holmes, P.:  {\em Celestial Encounters: The Origins of Chaos and Stability}.  Princeton University Press, Princeton , 1996.

 \bibitem{Henon1964}
 H\'{e}non, M. and Heiles, C.: ``The applicability of the third integral of motion: some numerical experiments''. {\em  Astrophys. J.}, \textbf{69}:73 -- 79 (1964).

\bibitem{Valtonen2005}
Valtonen, M. and Karttunen, H.: {\em The three-body problem}. Cambridge University, Cambridge, 2005.

\bibitem{Polchinshi1998}
Polchinski, J.:  {\em String Theory}.  Cambridge University Press, Cambridge, 1998.

\bibitem{Heisenberg1927}
  Heisenberg, W.:`` {\"{U}ber den anschaulichen Inhalt der quantentheoretischen Kinematik und Mechanik}'', {\em Zeitschrift f\"{u}r Physik}, \textbf{43}  (3-4): 172 --198 (1927).

\bibitem{Broglie1924}
de Broglie, L.:  Recherches sur la th\`{e}orie des quanta (Researches on the quantum theory), Thesis, Paris, 1924.

\bibitem{Sprott2010}
Sprott, J.C.: {\em Elegant Chaos}.  World Scientific, New Jersey, 2010.

\bibitem{Bai1994}
Y.L. Bai,  F.J. Ke,  and  M.F. Xia,   ``Deterministically stochastic behavior and sensitivity to initial configuration in damage fracture''.  {\em Science Bulletin}  \textbf{39}:  892Ð895 (1994).

\bibitem{Xia2000}
M.F.  Xia,  F.J. Ke, Y.J. Wei, J. Bai,  and  Y.L. Bai, ``Evolution induced catastrophe in a nonlinear dynamical model of material failure''. {\em Nonlinear Dynamics} \textbf{22}: 205 -- 224 (2000).

\bibitem{Sussman1988}
Sussman,~G.J.  and  Wisdom,~J.:  ``Numerical Evidence that the Motion of Pluto is Chaotic'',  {\em Science},  {\bf 241}:  433 -- 437 (1988).

\bibitem{Sussman1992}
Sussman,~G.J.  and  Wisdom,~J.: ``Chaotic Evolution of the Solar System'',  {\em Science}, {\bf 257}: 56 -- 62 (1992).

\bibitem{3bodyPRL-2013}
S\v{u}vakov, M.  and Dmitras\v{i}novi\'{c},~V.:  ``Three Classes of Newtonian Three-Body Planar Periodic Orbits'', {\em Phys. Rev. Lett.}, {\bf 110}:  114301 (2013).

\bibitem{quantum}
Griffiths, D. J.: {\em Introduction to Quantum Mechanics (2nd ed.)}. Prentice Hall (2004).

\end{thebibliography}
\end{document}